\documentclass[a4paper,fleqn,usenatbib,twocolumn]{mnras}

\usepackage[T1]{fontenc}
\usepackage{ae,aecompl}
\usepackage{graphicx}	
\usepackage{amsmath}	
\usepackage{amssymb}	
\usepackage[section]{placeins}

\title[Manifold spirals and secular evolution in galaxies]
{Manifold spirals, disc-halo interactions and the secular evolution 
in N-body models of barred galaxies}

\author[C. Efthymiopoulos et al.]{\Large
C. Efthymiopoulos$^{1}$\thanks{E-mail: cefthim@academyofathens.gr}
P.E. Kyziropoulos,$^{2}$\thanks{E-mail: pkyzirop@ee.duth.gr}
R.I. P\'aez$^{3}$\thanks{E-mail: paez@math.unipd.it}
K.Zouloumi$^{1,4}$\thanks{E-mail: konstantina-z7@hotmail.com}
and G.A. Gravvanis$^{2}$\thanks{E-mail: ggravvan@ee.duth.gr}
\\
$^{1}$Research Center for Astronomy, Academy of Athens, 
Soranou Efessiou 4, 11527 Athens, Greece\\
$^{2}$Department of Electrical and Computer Engineering, 
School of Engineering,\\
  Democritus University of Thrace, University Campus, 
Kimmeria, 67100 Xanthi, Greece\\
$^{3}$Department of Mathematics, University of Padova, 
Via Trieste 63, 35121 Padova, Italy\\
$^{4}$Department of Physics, University of Athens, 
Panepistimiopolis, 11521 Athens, Greece\\
}

\date{Accepted XXX. Received YYY; in original form ZZZ}

\pubyear{2019}

\begin{document}
\label{firstpage}
\pagerange{\pageref{firstpage}--\pageref{lastpage}}
\maketitle

\begin{abstract}
{\small The manifold theory of barred-spiral structure provides a dynamical 
mechanism explaining how spiral arms beyond the ends of galactic bars can be 
supported by chaotic flows extending beyond the bar's co-rotation zone. 
We discuss its applicability to N-body simulations of secularly evolving 
barred galaxies. In these simulations, we observe consecutive `incidents' of 
spiral activity, leading to a time-varying disc morphology. Besides disc 
self-excitations, we provide evidence of a newly noted excitation mechanism 
related to the `off-centering' effect: particles ejected in elongated orbits 
at major incidents cause the disc center-of-mass to recoil and be set in 
a wobble-type orbit with respect to the halo center of mass. The 
time-dependent $m=1$ perturbation on the disc by the above mechanism 
correlates with the excitation of new incidents of non-axisymmetric activity 
beyond the bar. At every new excitation, the manifolds act as dynamical 
avenues attracting particles which are directed far from corotation along 
chaotic orbits. The fact that the manifolds evolve morphologically in time, 
due to varying non-axisymmetric perturbations, allows to reconcile manifolds 
with the presence of multiple patterns and frequencies in the disc. We find 
a time-oscillating pattern speed profile $\Omega_p(R)$ at distances $R$ 
between the bar's corotation, at resonance with the succession of minima 
and maxima of the non-axisymmetric activity beyond the bar. Finally, we 
discuss disc thermalization, i.e., the evolution of the disc velocity 
dispersion profile and its connection with disc responsiveness to manifold 
spirals. }
\end{abstract}


\section{Introduction}
The manifold theory of spiral structure (\cite{rometal2006}; 
\cite{vogetal2006a}; see also \cite{dan1965}) provides a dynamical 
mechanism to interpret how structures such as the bi-symmetrical 
spiral arms, observed beyond the ends of a galactic bar, can arise out of 
chaotic motions of the underlying distribution of matter. 
Two versions of the manifold theory have been developed so far in the 
literature. 

i) In the `flux-tube' version (\cite{rometal2006}; 
\cite{rometal2007}; \cite{athetal2009a}; \cite{athetal2009b}; \cite{ath2010}; 
\cite{ath2012}), one considers continuous-in-time orbits describing 
the flow of matter away from the bar's unstable Lagrangian points L$_1$ 
and L$_2$, as viewed in a frame of reference rotating with angular speed 
equal to the bar's pattern speed $\Omega_p$. The 
unstable `flux-tube' manifolds are invariant sets formed by all orbits 
tending asymptotically towards $L_1$ or $L_2$ in the backward sense of 
time. This means that, in the forward sense of time, these orbits form 
an outflow directed away from $L_1$ or $L_2$. An elementary linearization 
of the equations of motion around $L_1$ or $L_2$ shows that these outflows, 
or `flux-tubes', have the form of trailing spiral arms. Besides $L_1$ and 
$L_2$, similar flux-tube manifolds can be constructed for the whole family 
of epicyclic periodic orbits (called `Lyapunov orbits') around $L_1$ or 
$L_2$. 

ii) In the version of the manifold theory called `apocentric 
manifolds' (\cite{vogetal2006a}; \cite{tsouetal2008}; \cite{tsouetal2009}; 
\cite{eft2010}; \cite{haretal2016}) one computes first the flux-tube 
manifolds, and then isolates only those points which correspond to 
apsidal positions, i.e., local apocentric or pericentric points of 
each orbit in the flux-tube. One can show that the pattern formed by 
the union of the apocentric points yields again trailing spiral arms. 
Near $L_1$ or $L_2$ the shapes of the flux-tube and apocentric manifolds 
coincide. However, far from the Lagrangian points, the shapes of the 
apocentric manifolds allow to visualize the intricate chaotic dynamics 
known in dynamical systems' terminology as the `homoclinic tangle' 
(see \cite{wig1990}). Thus, the flux-tube and the apocentric manifolds 
are the same phase-space objects, but visualized differently in physical 
space. For a detailed comparison and generalizations of the various 
manifold theories see \cite{haretal2016}, or the tutorial \cite{eft2010}. 

Since the manifold spirals arise in a system of reference which co-rotates 
with the bar, the manifold theory in its basic form predicts that the 
spiral arms should have the same pattern speed as the bar. This remark 
seems to come in conflict with observations both in our Galaxy (as reviewed 
e.g. in \cite{blager2016}; see also \cite{antetal2014}, \cite{junetal2015} 
and references therein) and in other galaxies (e.g. \cite{veretal2001}; 
\cite{booetal2005}; \cite{patetal2009}; \cite{meietal2009}; \cite{spewes2012}; 
\cite{speroo2016}). Regarding, now, galactic disc simulations, the leading 
paradigm over the years refers to simulations showing the co-existence of 
multiple pattern speeds (\cite{selspa1988}; \cite{litcar1991}; \cite{rausal1999}; 
\cite{qui2003}; \cite{minqui2006}; \cite{dubetal2009}; \cite{quietal2011}; 
\cite{minetal2012}; \cite{babetal2013}; \cite{rocetal2013}; \cite{fonetal2014}; 
\cite{bab2015}; but see also a noticeable exception in \cite{rocetal2013}), 
possibly connected also to the phenomenon of nonlinear coupling of multiple  
disc modes (\cite{tagetal1987}; \cite{tagath1991}; \cite{selwil1993}; 
\cite{mastag1997}). On the other hand, it is well known that even isolated 
barred galaxies undergo substantial secular evolution (see \cite{ath2013}; 
\cite{bin2013}; \cite{kor2013} in the tutorial volume \cite{falkna2013}). 
The tendency to transfer angular momentum outwards (e.g. towards the halo 
or across the disc, \cite{trewei1984}; \cite{debsel1998}; \cite{debsel2000}; 
\cite{ath2002};  \cite{athmis2002}; \cite{ath2003}; \cite{oneidub2003}; 
\cite{holbocketal2005}; \cite{beretal2006}; \cite{marvaletal2006}) leads the 
bar to slow down and grow in size at a rate which produces non-negligible change 
in dynamics at timescales comparable even to a few bar periods. This process 
becomes complex, and even partially reversed due to the growth of 
`pseudo-bulges' or peanuts (\cite{korken2004}), caused by dynamical 
instabilities such as chaos or the `buckling instability' \cite{comsan1981}; 
\cite{cometal1990}; \cite{pfefri1991}; \cite{rahetal1991}; \cite{burath1999}; 
\cite{marshl2004}; \cite{burath2005}; \cite{debetal2006}). The reduction in 
size of the bar by the transfer of angular momentum under constant pattern 
speed is discussed in \cite{weikat2007}. Spiral activity acts as an 
additional factor of outwards transfer of angular momentum (\cite{lynkal1972}), 
while a radial re-distribution of matter can take place even under a 
nearly-preserved distribution of angular momentum (\cite{hohl1971}; 
\cite{selbin2002}; \cite{avietal2005}). Radial migration is enhanced by 
the amplification of chaos due to the overlapping of resonances among the 
various patterns (\cite{qui2003}; \cite{minqui2006}; \cite{quietal2011}). 

As secular evolution affects both the bar's morphology and pattern speed, 
as well as the extent to which other non-axisymmetric patterns are present 
in the disc, the shapes of the invariant manifolds found by momentarily 
`freezing' the potential and pattern speed value also undergo important 
changes in time. Simulations by Lia Athanassoula have shown 
that, despite these changes, the stars out-flowing from the neighborhood 
of the Lagrangian points $L_1$ and $L_2$ develop orbits which, in general 
keep track of the change of the form of the invariant manifolds 
(\cite{ath2012}; see also \cite{bab2015}; \cite{lok2016}). As a rule, 
the material which populates the manifolds comes from orbital outflows 
originating from the interior of co-rotation, at the end of the bar 
(\cite{con1980}). As these outflows are adapted to the slowly-changing 
form of the manifolds, they are able to yield time-varying spiral or 
ring-like patterns. In principle, this non-rigidity of the shape and 
population density of the invariant manifolds leaves space to reconcile 
manifold spirals with time and space varying pattern speeds beyond the bar, 
although such a reconciliation awaits probe by specific simulations. 

In the present paper we aim to investigate the application of the
manifold theory of spirals in a typical model of secularly evolving
bar. To this end, a series of isolated disc galaxy simulations were
performed using the explicit Mesh-Adaptive N-Body method based on
Approximate Inverses (MAIN, \cite{kyzetal2016}), and the Parallel
Self-Mesh Adaptive N-body technique based on Approximate Inverses
(PMAIN, \cite{kyzetal2017a}; \cite{kyzetal2017b}).  The initial
conditions for these simulations are described in \cite{kyzetal2016}.
Whenever a bar is formed in a simulation, we find manifold spirals, as
well as traces of secular evolution affecting the bar and other
  non-axisymmetric features in the disc. In the present paper we
focus on one of these simulations (`Q1' in \cite{kyzetal2016}), aiming
to present these phenomena in considerable detail.  This is a typical
simulation starting with a Toomre $Q=1$ exponential disc, a
Dehnen-like `double-power-law' halo, and a Sersic bulge, with
parameters corresponding to a Milky-Way type galaxy (see \cite{bintre2008}).  
The simulation undergoes phases of evolution known in literature since 
decades (see references above), namely: i) an initial instability leading 
to the formation of $m=2$ spirals, lasting for $\sim 1Gyr$, ii) a bar 
instability, erasing all traces of the previous spirals and generating 
new ones related to the bar, iii) a `bucking instability' which thickens 
the bar, and iv) a phase of slow evolution in which the amplitude of 
spirals is reduced.  We focus our study on the epoch after the bar is 
formed (around $t=1.5$Gyr), and up to $t=4Gyr$. We measure several 
quantities which quantify the bar's secular evolution. We then observe 
the corresponding evolution of the manifolds emanating from the
neighborhood of $L_1$ and $L_2$, as well as what happens, in general, 
to the disc as the bar-spiral structure evolves.

Our results lead to the following picture: as typical in such
simulations, we find that spiral activity beyond the bar is
characterized by recurrent episodes. `Incidents' of growth of
spiral and other non-axisymmetric disc features with substantial
spectral power take place in a nearly damped-oscillatory manner.
Damping, which results in these non-axisymmetric features gradually
fading in amplitude, can be attributed to `disc heating', i.e. the
overall increase in time of the velocity dispersion in the outer parts
of the disc (see \cite{sel2014} for a review). However, regarding what
causes, in the first place, the repeated excitations of incidents of
non-axisymmetric activity, besides recognizing in our simulation
some well known mechanisms of enhancement of local perturbations
(e.g. swing amplification, see also \cite{bab2015}), we find evidence
of a mechanism not emphasized, to our knowledge, in literature.  This
is the fact that, every incident is accompanied by the creation of a
population of particles moving in orbits either escaping or with 
apocenters far from the main part of the disc, whose distribution of 
directions of motion exhibits fluctuations with respect to perfect 
central symmetry. Hence, they cause the main part of the disc to recoil 
with respect to the spheroid (i.e. halo-bulge) center of mass. The  
recoil causes, in turn, an `off-centering effect' resulting in a $m=1$ 
disturbance on the disc. Some studies have indicated the importance of 
such phenomena in the evolution  of bar-halo systems, as well as the 
evolution of halo cusps (\cite{sel2003}; \cite{mcmdeh2005}; 
\cite{holbocketal2005}; \cite{deb2006}; see also \cite{hametal2018} for 
a different application in the relaxation of globular clusters). Here 
we find that the `incidents' of non-axisymmetric activity are correlated 
with the off-centering effect. The disc recoil sets the disc center of 
mass in relative orbit with respect to the spheroid (halo-bulge) center 
of mass. The relative orbit has small size (the `wobbling' between disc 
and halo can extend as much as $\sim 0.3$ kpc, about $5-10\%$ of the bar 
size; off-centering effects of larger size are reported in \cite{mcmdeh2005}, 
with no apparent connection with the mechanism mentioned above). Yet, this
is sufficient to induce a significant $m=1$ perturbation on the disc,
with amplitude $\sim 0.1$ in a region including the bar. The disc response 
is found to have both $m=1$ and $m=2$ components. The latter might be 
connected to particles pushed to regions outside co-rotation, in which 
case they have to follow chaotic orbits guided by the (predominantly $m=2$) 
manifolds. Since at least one major recoil event is observed before the 
bar, the manifolds cannot be the drivers of the phenomenon. They can, 
however, help to maintain it, by supporting particles in chaotic orbits 
leading to fluctuations in the directions in which the particles move.

All together, we identify the role of the invariant manifolds as a skeleton 
of chaotic orbits in phase space, or, the dynamical avenues to be followed 
by particles whenever an incident of spiral or other non-axisymmetric is 
triggered in the disc. The morphology of observed patterns in the disc is 
always qualitatively similar to the `chaotic tangle' formed by the manifolds, 
while direct comparison of the two patterns yields various levels of 
agreement, depending on the snapshot examined.

We finally study how all these phenomena evolve with the change in disc 
`temperature' (velocity dispersion). The incidents of non-axisymmetric 
activity beyond the bar fade in a timescale $~4$ Gyr. However, this 
happens in a purely stellar dynamical simulation. Invoking mechanisms 
able to dissipate kinetic energy in random motions (e.g. gas cooling, 
star formation etc.)  should help to prolong these phenomena over a 
considerable fraction of the age of real disc galaxies.

The paper is structured as follows: section 2 presents the N-body
simulation as well as methods of analysis. Section 3 explains the
computation of the invariant manifolds and its comparison with the
disc morphology. Section 4 deals with the disc's secular evolution, 
and shows the comparison with the manifolds at various times. Section 5 
summarizes our conclusions, along with some further comments on the role 
of manifolds in the modelling of barred-spiral galaxies.

\section{Simulation and numerical computations}

\subsection{Initial conditions and N-body simulation}
The simulation analyzed below corresponds to the experiment called `Q1' 
in \cite{kyzetal2016}. In summary, we use $10^7$ particles to simulate:

i) an initially exponential disc of mass $M_{disc}=5\times 10^{10} M_{\odot}$, 
exponential scale length $R_d=3$ kpc, vertical exponential scalelength 
$z_d=0.2$ kpc and Gaussian velocity distribution arising from a profile 
of Toomre's Q-parameter rising in the center and tending to an asymptotic 
outward value $Q\rightarrow Q_\infty$ with $Q_\infty=1$ (see equation (13) 
of \cite{kyzetal2016}). 

ii) a Sersic-type spherical bulge embedded in the disc, mass 
$M_b=5\times 10^9$, Sersic index $n=3.5$, scale length $R_b=1$ kpc. 

iii) A Dehnen-type double-power law spherical dark matter halo  
(see \cite{bintre2008}), with density parameter $\rho_0=2.016\times 10^8$ 
$M_{\odot}/kpc^3$, scale length $R_h=3$ kpc, and asymptotic inner and outer 
density power-law exponents $\alpha=1.3$, $\beta =3.5$. The halo yields 
a mass $\approx 7\times 10^{10}M_{\odot}$ at $R=50$ kpc, which rises 
to $2\times 10^{11}M_{\odot}$ at $R=100$ kpc.

The disc, bulge and halo are represented by $5\times 10^6$, $5\times 10^5$ 
and $4.5\times 10^6$ particles respectively. A `relaxation technique' is used 
to match all three components in a unique system with virial ratio very close 
to $2$ (see \cite{kyzetal2016} and \cite{kyzetal2017a} for a description).  
Being prone to disc instabilities, the system exhibits vivid evolution dominated 
by the succession of spiral, bar, and other non-axisymmetric features.

The simulation was performed using the N-body code MAIN, which
utilizes a Cartesian grid that solves Poisson equation by the use of a
novel Symmetric Factored Approximate Sparse Inverse (SFASI)pre-conditioning 
technique, in conjunction with the multigrid method. Details on the features 
of the numerical technique are provided in \cite{kyzetal2016}. In 
\cite{kyzetal2017a} and \cite{kyzetal2017b} we presented two fully parallel 
versions of the N-body technique (for shared-memory and distributed-memory 
systems respectively).  These versions implement fully adaptive mesh
refinement while improving the algorithm by which boundary conditions
are computed on the sides of the computational box. Using these improved 
techniques, we performed $N=10^8$ and $N=10^9$ particle versions of the 
numerical simulation Q1 (see \cite{kyzetal2017a} and \cite{kyzetal2017b}). 
The new simulations differ from the original one essentially only with 
respect to the exact moment when the bar instability is manifested. This 
is partly due to the better resolution of the profile of the central 
force, which affects the onset of the bar instability (\cite{holbocketal2005}), 
and also to the fact that a smoother particle distribution implies 
lesser noise, and hence longer time for any instabilities to grow in the 
disc (e.g. by swing amplification, \cite{tookal1991}). On the other hand, 
the main features of the bar and spirals (Fourier amplitudes, pattern speeds, 
secular evolution etc) appear similar in all runs.

Several computations require use of a smooth approximation of the potential 
$\Phi(x,y,t)$ in the disc plane $(x,y)$. This is obtained, using bi-cubic 
interpolation, from the N-body potential evaluated in a grid $x_i=-L_0/2 + 
(i/N_p)L_0$, $y_j=L_0/2 + (j/N_p)L_0$, where $i,j=0,1, \ldots,N_p-1$, 
$L_0=25.6$ kpc, $N_p=256$.

\subsection{Evolution of non-axisymmetric patterns}
\begin{figure}
\centering
\includegraphics[scale=0.26]{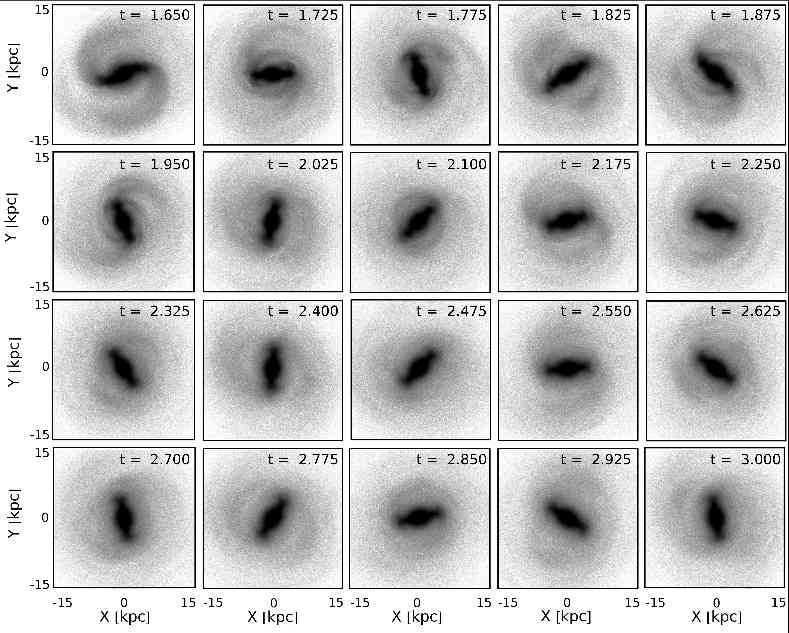}
\caption{\small Twenty snapshots of the face-on disc view in the
interval $1.65\leq t[Gyr] \leq 3.0$. The bar rotates counter-clockwise. 
Spiral and ring-like structures appear reccurently and with varying 
amplitudes.}
\label{fig:rabdosplots2}
\end{figure}

As a consequence of the chosen initial value of $Q$ (=1), the disc 
exhibits a rapid growth of competing features. A $m=2$ spiral mode 
dominates up to about $t=1.1$ Gyr. After some transient phase, the bar 
starts growing rapidly between $t=1.3$ Gyr and $t=1.6$ Gyr. This growth is 
followed by `incidents' of spiral activity (see below). Five such `incidents' 
take place up to $t=2$ Gyr. The overall morphology of the galaxy changes 
rapidly in timescales as small as the bar period. The morphological evolution 
of the disc is shown in Fig. \ref{fig:rabdosplots2}. As typical in such 
simulations, features like rings and spiral arms appear reccurently and 
with varying morphology after the bar formation. We focus on the incidents 
of spiral activity which take place between $t=1.65$ and $t=3$ Gyr. The bar 
performs slightly more than seven (counterclockwise) revolutions in this time 
interval. The spiral modes oscillate from conspicuous maxima (e.g. at 
$t=1.65$, $t=1.95$, $t=2.25$ Gyr) to very low minima 
(e.g. at $t=2.025$, $2.325$, $2.7$, $2.925$ Gyr), albeit never vanishing 
completely. An inner ring-like structure surrounding the bar is also 
present, while, at particular snapshots (e.g. $t=1.825$) material appears 
to travel from one end of the bar to the other along the ring. This 
phenomenon is discussed in depth in several papers on the `flux-tube' 
manifolds, e.g. \cite{rometal2006}; \cite{ath2012}. 

The bar also thickens in time, forming a peanut, or `X-shaped', pseudobulge). 
From an inspection of the vertical disc profiles and the evolution of the $m=2$ amplitude (see below), we infer that a buckling instability occurs around 
$t\approx 2$ to $2.1$ Gyr. Manifold spirals, however, keep appearing well 
after $t=2.1$ Gyr. Thus, the manifold spirals do not appear to be halted 
by the buckling instability, as was reported for other simulations in 
\cite{kwaetal2017} (see also \cite{lok2016}).
\begin{figure}
\begin{center}
\includegraphics[scale=0.24,angle=0]{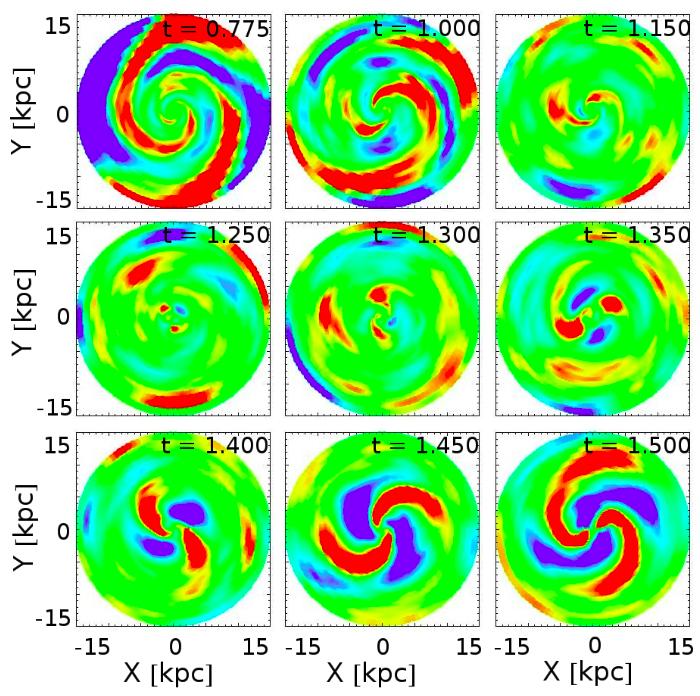}
\end{center}
\caption{\small The non-axisymmetric excess density $D(R,\phi)$ in 
nine snapshots in the interval $0.775\leq t[Gyr]\leq 1.5$. The color scale 
represents values $D(R,\phi)\leq -0.5$ in blue, $D(R,\phi)\geq 0.5$ 
in red and $-0.5<D(R,\phi)<0.5$ in between. The transition from a spiral 
to a bar mode dominance is shown. The first manifold spirals are developed 
immediately after the bar formation.}
\label{fig:fourierbar}
\end{figure}
\begin{figure}
\begin{center}
\includegraphics[scale=0.25,angle=0]{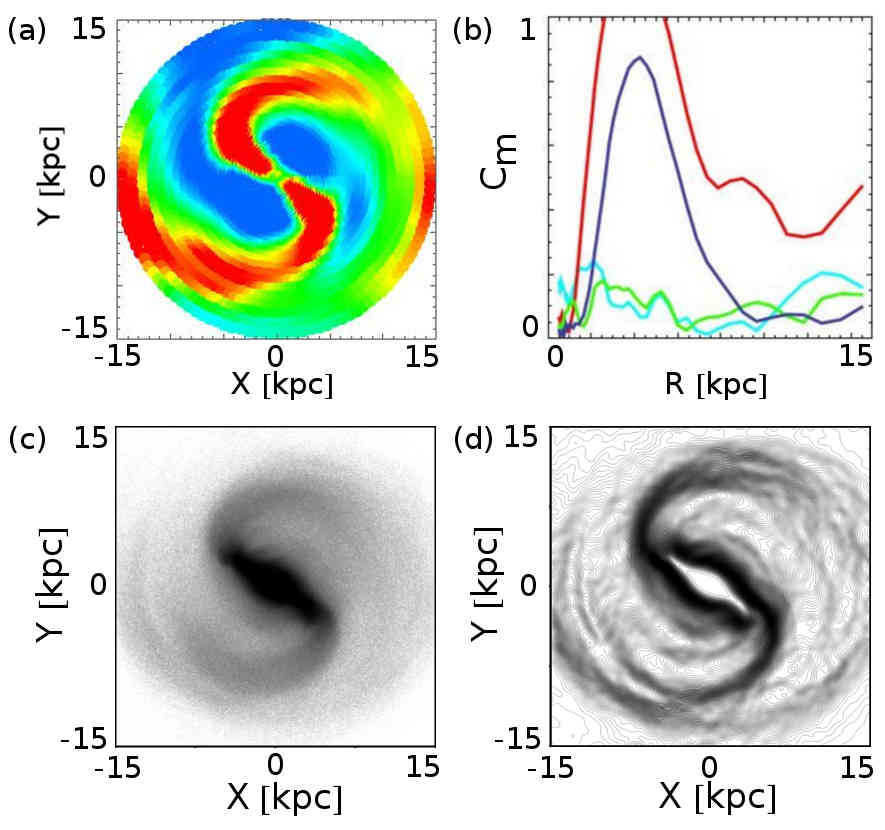}
\end{center}
\caption{\small (a) Color map of the non-axisymmetric excess density 
at $t=1.625$ Gyr. A second incident of spiral activity accompanies the bar. 
(b) The Fourier amplitudes $C_1(R)$ (cyan), $C_2(R)$ (red), $C_3(R)$ (green) 
and $C_4(R)$ (blue). (c) The disc image by a logarithmic grayscale plot of 
the surface density $\Sigma(R,\phi)$. (d) The same image as in (c) after 
implementing the Sobel-Feldman edge detection algorithm (see text). Dimmer 
patterns of (c) are enhanced and clearly visible in (d). }
\label{fig:snap65}
\end{figure}
We quantify the evolution of non-axisymmetric features using the disc 
surface density
\begin{equation}\label{eq:sig}
\Sigma_d(R,\phi,t) = {\Delta N(R,\phi,t)\over R\Delta R\Delta\phi} 
\end{equation}
where $\Delta N(R,\phi,t)$ is the number of disc particles at time $t$ in 
each area element $R\Delta R\Delta\phi$ of a polar grid of 50 logarithmically 
equi-spaced radial bins from $R_1=0.1$ kpc to $R=15$ kpc, and 180 azimuthal 
bins from $\phi=0$ to $\phi=2\pi$.
Next, we compute the Fourier transform
\begin{equation}\label{eq:foursig}
\Sigma_d(R,\phi)=\Sigma_{d,0}(R) 
+ \sum_{m=1}^\infty [A_m(R)\cos(m\phi)+B_m(R)\sin(m\phi)]~.
\end{equation}
The relative amplitude $C_m(R)$ and phase $\phi_m(R)$ of the m-th Fourier 
mode are defined by
\begin{equation}\label{eq:ampphase}
C_m = {\left(A_m^2+B_m^2\right)^{1/2} \over \Sigma_{d,0}},~
\phi_m = {1\over m}\tan^{-1}\left({B_m\over A_m}\right)~.
\end{equation}
Truncating Eq.~(\ref{eq:foursig}) in a finite number of harmonics
($m\leq 10$) yields a smoothed representation of the surface density
$\Sigma_s(R,\phi)$. The smoothed `non-axisymmetric excess density'
is defined as
\begin{equation}\label{eq:exden}
D(R,\phi) = {\Sigma_s(R,\phi)-\Sigma_{d,0}(R) 
\over \Sigma_{d,0}}~~~. 
\end{equation}
The rapid growth of non-axisymmetric features in the disc leads 
to a dominant $m=2$ spiral mode lasting up to the time $t = 1.1$ Gyr
(Fig. \ref{fig:fourierbar}). However, the onset of the bar instability 
erases all traces of previous spiral activity. The bar grows rapidly 
between $t=1.3$ Gyr and $t=1.6$ Gyr. A first conspicuous set of 
spiral arms beyond the bar appears around $t=1.5$ Gyr. The bar extends 
to $R\approx 5$ kpc (as measured by the interval in $R$ in which the 
$m=2$ phase remains nearly constant), while co-rotation is initially 
at $R\approx$ 6.5kpc. At $t=1.625$ Gyr, a second major incident of 
spiral activity takes place. Beyond this time, the $m=2$ and $m=4$ 
Fourier amplitudes enter a phase of considerably slower evolution.

Setting conventionally $t=1.625$ as the time at which the bar's
secular evolution begins, in the next section we focus on this time
snapshot to describe the computations related to manifold spirals
and the comparison between manifolds and observed N-body disc
morphologies. As shown in Fig.\ref{fig:snap65}, the bar is very
strong at this snapshot, with $C_2>1$ and $C_4\approx 0.8$ at
$R\approx 4$ kpc. There is also important power in the $m=1$ and
$m=3$ Fourier terms. The $C_1$ amplitude is larger than 0.1 across 
the bar's whole extent and reaches $\sim 0.2$ in the central parts 
($R<2$ kpc). Such $m=1$ amplitudes are identifiable in
several simulations (see, for example, \cite{quietal2011};
\cite{minetal2012}). The question of what triggers the $m=1$ mode in
the inner parts of the disc will be dealt with in section 4.  On the
other hand, in the snapshot of Fig. \ref{fig:snap65}, substantial
$m=1$ relative amplitudes are observed also at large distances ($R>10$
kpc). This is connected to the lopesidedness of the observed
spirals (and manifolds, see below), a fact which hints towards
nonlinear interaction of the outer disc modes.

In Fig. \ref{fig:snap65}, the density excess map (a) clearly shows the main 
spiral pattern, but not so clearly secondary features as the interarm features 
or the bifurcation of spirals seen in the lower left part of the direct image 
of the disc (panel (c)). In order to bring out these features, we produce an 
image of the disc (panel (d)) which enhances the edges of most `eye-recognizable'
patterns of image (c) using a pattern-recognition algorithm known as the 
`Sobel-Feldman edge detection'(see \cite{gonwoo1992}). This technique 
is similar to unsharp masking in that it processes the image through a 
smoothing filter (we used a Gausssian filter for image (c)). Instead of 
substracting the two images, Sobel-Feldman uses an operator to approximate 
the gradient of the smoothed image, and thus detect the edges of patterns, 
where the gradient becomes large. The Sobel-Feldman technique enhances the 
ability to recognize non-axisymmetric features of amplitude significantly 
lower than the one of the main bi-symmetric spirals. In subsequent sections 
we will use such `Sobel-Ferldman' images of the disc in comparison with the 
patterns formed by the manifolds in the corresponding snapshots.

\section{Manifold spirals}

\subsection{Definitions}
At a fixed time $t$, consider observers moving in a galactocentric frame 
rotating with angular velocity $\Omega_p$ equal to the instantaneous angular 
velocity of the bar. Let $V(x,y,z)$ be the total gravitational potential 
(arising from all components, e.g. bulge, disc-bar, halo) at the time $t$. 
We consider first orbits in the rotating frame with `frozen' in time potential 
$V$. Let $\Phi(x,y)=V(x,y,z=0)$ be the potential restricted in the disc plane. 
The orbits are given by Hamilton's equations with Hamiltonian
\begin{equation}\label{eq:ham2d}
H = {p_R^2\over 2} + {p_\phi^2\over 2R^2} -\Omega_p p_\phi +\Phi(R,\phi)~~,
\end{equation}
where $(R,\phi)$ are cylindrical co-ordinates in the rotating frame, 
$x=R\cos\phi$, $y=R\sin\phi$, and $(p_R,p_\phi)$ are the radial and angular 
momentum (per unit mass) in the intertial frame, connected to the velocities 
$V_x,V_y$ in the rotating frame via $p_R=(xV_x+yV_y)/R$, $p_\phi = 
\Omega_pR^2+(xV_y-yV_x)$. The potential $\Phi(R,\phi)$ admits the Fourier 
decomposition:
\begin{equation}\label{eq:potfour}
\Phi(R,\phi) = \Phi_0(R) + \sum_{m=1}^\infty[\Phi_m(R)\cos(m\phi) 
+ \Psi_m(R)\sin(\phi)]~.
\end{equation}

Consider first a simplified bar-like model in which $\Phi_m(R) = 
\Psi_m(R)=0$ for all $m>0$ except $m=2$ and $\Phi_2(R)/\Psi_2(R)=const$. 
The bar has a fixed orientation angle $\phi_{bar}$ in the rotating frame. 
The equations of motion yield four Lagrangian equilibrium points 
(\cite{bintre2008}). The unstable Lagrangian points are at the end of 
the bar, $R_{L1} = R_{L2}=R_C$, $p_{\phi,L1}=p_{\phi,L2}=\Omega_p R_C^2$. 
The stable Lagrangian points are at distance $R_C'$. The annulus 
$R_C'<R<R_C$ is hereafter called the corotation zone. Introducing the 
full non-axisymmetric perturbation may induce an important deformation 
of the equipotential surfaces, influencing the number and position of 
unstable Lagrangian points (\cite{tsouetal2009}, \cite{athetal2009a},
\cite{athetal2009b}, \cite{kaletal2010}, \cite{wuetal2016}). In the 
present simulation we observe only a small displacement of the points 
$L_1$, $L_2$ with respect to the pure bar model.

The key property leading to the definition of the manifold spirals is 
now the following: one can show that there is a  family 
of initial conditions for which the resulting orbits under the potential 
(\ref{eq:potfour}) tend asymptotically to the Lagrangian point $L_1$ when  
integrated backward in time, i.e., as $t\rightarrow-\infty$. This leads to 
the formal definition of the {\it unstable manifold} of $L_1$. Denote by 
$O(t;R_0,\phi_0,p_{R,0},p_{\phi,0})$ one orbit with initial conditions 
$(R_0,\phi_0,p_{R,0},p_{\phi,0})$, and also $O_{L1}\equiv(R_{L1},\phi_{L1},
p_{R,L1},p_{\phi,L1})$. The unstable manifold ${\cal W}^U_{L1}$ is the 
ensemble of all different initial conditions for which the distance between 
$O(t;R_0,\phi_0,p_{R,0},p_{\phi,0})$ and $O_{L1}$ tends to zero as 
$t\rightarrow -\infty$:
\begin{eqnarray}\label{eq:wul1}
  {\cal W}^U_{L1} = \Bigg\{\mbox{All~}
  (R_0,\phi_0,p_{R,0},p_{\phi,0}):
~~~~~~~~~~~~~~~~~~~~~~~\nonumber \\
 dist[O(t;R_0,\phi_0,p_{R,0},p_{\phi,0}),O_{L1}]\rightarrow 0 
\mbox{~as~} t\rightarrow -\infty\Bigg\}~.
\end{eqnarray}
Similar definitions hold for the unstable manifold of the Lagrangian point 
$L_2$. Basic theorems of dynamics (\cite{gro1959}; \cite{har1960}) ensure 
that the sets ${\cal W}^U_{L1}$, ${\cal W}^U_{L2}$ are non-empty. 
The dynamical role of the invariant manifolds ${\cal W}^U_{L1}$ and 
${\cal W}^U_{L2}$ can be summarized as follows: ${\cal W}^U_{L1}$ collects 
the ensemble of trajectories tending arbitrarily close to $L_1$ in the 
backward sense of time. Thus, starting very close to $L_1$, and integrating 
{\it forward} in time, these trajectories form an outflow away from $L_1$. 
Projected in real space, this outflow yields the `flux-tube' manifold 
${\cal W}^{UFT}_{L1}$ (\cite{rometal2006}; \cite{rometal2007}). An elementary 
analysis  shows that the flux-tube manifold has two branches: one is directed 
outwards (outside co-rotation), and it takes the shape of a trailing spiral 
arm. The second is directed inwards (inside corotation), creating a ring-like 
structure around the bar. Thus, the flux-tube manifolds can give rise to 
several morphological structures, from rings to open spirals, depending on 
the model's parameters (e.g. pattern spead, $m=2$ amplitude and asymmetry) 
as discussed in (\cite{athetal2009a}). 

In a similar way as for ${\cal W}^U_{L1}$, we can define outflows of 
asymptotic trajectories emanating from a small but finite distance from 
$L_1$. To this end, we invoke the family of short period (or `Lyapunov') 
unstable periodic orbits which form small retrograde epicycles around 
$L_1$, called hereafter the `PL1 orbits' (\cite{vogetal2006a}) (denoted 
$O_{PL1}$). Similarly to $L_1$, the unstable manifold of such an orbit, 
is defined as 
\begin{eqnarray}\label{eq:wupl1}
{\cal W}^U_{PL1} = \Bigg\{\mbox{All~} 
(R_0,\phi_0,p_{R,0},p_{\phi,0}): 
~~~~~~~~~~~~~~~~~~~~~~~~\nonumber\\
dist[O(t;R_0,\phi_0,p_{R,0},p_{\phi,0}),O_{PL1}]\rightarrow 0 
\mbox{~as~} t\rightarrow -\infty\Bigg\}~.
\end{eqnarray}
The projection of the manifolds ${\cal W}^U_{PL1,2}$ in the disc plane 
yields the `flux-tube' manifolds ${\cal W}^{UFT}_{PL1,2}$. The geometric shapes 
and properties of the manifolds ${\cal W}^{UFT}_{PL1,2}$ are similar to those 
of the manifolds ${\cal W}^{UFT}_{L1,2}$. However, the 
periodic orbits PL1,2 and their manifolds form families which span a whole set 
of values of the Jacobi constant $E_J$, while the unstable points $L_{1,2}$ 
and their manifolds represent only the values of the Jacobi constants 
$E_{J,L1,2}$. One has $E_{J,PL1,2}>E_{J,L1,2}$ for any orbit of the PL1,2 
families and their manifolds, implying that the manifolds ${\cal W}^{UFT}_{PL1,2}$ 
characterize the motions of particles at all energies beyond the Jacobi energy 
at corotation.  

The `flux-tube' manifolds defined above represent a continuous swarm 
of trajectories with initial conditions in the set (\ref{eq:wul1}), or 
(\ref{eq:wupl1}). This set reproduces the manifolds as geometric objects, 
but provides no information on the real distribution of matter along the 
manifolds, which depends on the distribution function of the N-body system. 
Despite the fact that a precise knowledge of the distribution function is 
hardly tractable, the method of apocentric manifolds (\cite{vogetal2006a}) 
exploits the fact that local density maxima along the manifolds are expected 
at points close to apsidal (i.e. pericentric or apocentric) positions of the 
orbits. This assumption is verified in N-body experiments (see 
\cite{tsouetal2008}), where it is shown that density maxima corresponding, 
in particular, to spiral patterns are associated with local apocenters of 
the orbits of the N-body particles.  The apocentric manifolds 
${\cal W}^{UA}_{L1,2}$ are defined through the manifolds ${\cal W}^{U}_{L1,2}$ 
as follows:
\begin{equation}\label{eq:wuapo}
{\cal W}^{UA}_{L1,2} = 
\Bigg\{
\mbox{All points of~}{\cal W}^{U}_{L1,2}: p_R=0,\dot{p}_R<0
\Bigg\}~~.
\end{equation}
A similar definition holds for the apocentric manifolds of the families PL1,2.
For visualization purposes, the main benefit of the apocentric manifolds is that 
they allow to plot longer parts of the invariant manifolds, thus allowing to 
visualize in physical space the intricate chaotic dynamics induced by these 
manifolds, otherwise known as the `homoclinic' or `chaotic tangle' (see 
\cite{eft2010}). In the sequel we present computations based on the apocentric 
manifolds ${\cal W}^{UA}_{PL1,2}$. In practical steps, we compute first an 
`apocentric surface of section' for all trajectories of given Jacobi energy 
$E_J$. Setting $p_R=0$, $\dot{p}_R<0$, the equation 
\begin{equation}\label{eq:aposos}
H(R,\phi,p_R=0,p_\phi)=E_J 
\end{equation}
allows to compute the local radius $R$ at which a particle with
trajectory corresponding to the Jacobi energy $E_J$ reaches a local
apocentric passage with values of its angular variables equal to
$(\phi,p_\phi)$.  Eq.~(\ref{eq:aposos}) fixes the value of $R$ as
a function of $(\phi,p_\phi)$.  Thus, for a randomly chosen
trajectory undergoing epicyclic oscillations, the transition from one
to the next apocentric passage can be viewed as a mapping
\begin{equation}\label{eq:sosmap}
(\phi,p_\phi)\rightarrow
(\phi'=F(\phi,p_\phi),p_\phi'=G(\phi,p_\phi))
\end{equation}
with the functions $F,G$ specified numerically via the integration of the 
trajectories. The orbits $PL1,2$ correspond to fixed points of the mapping 
(\ref{eq:sosmap}), i.e., where the following condition is satisfied: 
\begin{equation}\label{eq:pl12}
F(\phi_0,p_{\phi,0})=\phi_0,~~~G(\phi_0,p_{\phi,0})=p_{\phi,0}~~~. 
\end{equation}
Eqs.~(\ref{eq:pl12}) can be viewed as a $2\times 2$ algebraic system which 
can be solved numerically, using a root-finding technique, in order to 
compute the initial conditions $(\phi_0,p_{\phi_0})$ of the corresponding 
PL1 or PL2 orbit. Numerical differentiation allows to compute also the 
elements of the monodromy matrix:
\begin{equation}\label{mon}
M = \left[
\begin{array}{cc}
{\partial F\over\partial\phi} &{\partial F\over\partial p_\phi} \\
{\partial G\over\partial\phi} &{\partial G\over\partial p_\phi} 
\end{array}
\right]_{\phi=\phi_0,p_\phi=p_{\phi,0}}~~.
\end{equation}
The matrix $M$ has two real eigenvalues satisfying $\lambda_1\lambda_2=1$. 
The linear eigenvectors associated with the absolutely largest eigenvalue 
define a slope, or `eigendirection', of the linearized apocentric mapping 
in the neighborhood of the fixed point $(\phi_0,p_{\phi,0})$. To compute 
and visualize the apocentric invariant manifolds ${\cal W}^{UA}_{PL1,2}$ 
one takes many initial conditions equi-spaced along the unstable 
eigendirection and belonging to a linear segment of small total length 
$dS$ on the apocentric surface of section $(\phi,p_\phi)$. Integrating all 
these trajectories forward in time yields the manifolds ${\cal W}^{U}_{PL1,2}$. 
Taking only the iterates of the mapping (\ref{eq:sosmap}) on the apocentric 
surface of section yields the apocentric manifolds ${\cal W}^{UA}_{PL1,2}$. 
Note that the iterates of the mapping (\ref{eq:sosmap}) are pairs of values 
$(\phi,p_{\phi})$. However, the constant energy condition 
$H(R,\phi,p_R=0,p_\phi)=E_J$  (Eq.~(\ref{eq:aposos})) allows to compute the 
apocentric radius $R$ for any pair $(\phi,p_{\phi})$. Thus, any point on the 
apocentric surface of section corresponds to a triplet of values 
$(R,\phi,p_\phi)$. The visualization  of the apocentric manifold in physical 
space is obtained by plotting the manifold's computed iterated points 
$x=R\cos\phi$, $y=R\sin\phi$. This is equivalent to computing the 
flux-tube manifolds ${\cal W}^{UFT}_{PL1,2}$ and keeping only the points 
where the orbits have apocenter. On the other hand, the visualization 
of the apocentric manifolds in phase space (phase portrait) is obtained 
by plotting the manifold's iterated points $(\phi,p_\phi)$.

\subsection{Manifold spirals in the N-body simulation}
The computation of the apocentric manifolds ${\cal W}^{UA}_{PL1,2}$ in
our N-body computation proceeds in the general way described above.
Besides interpolating the N-body disc potential (see section 2), in
order to compute the equations of motion, we first specify the value
of the bar pattern speed as follows: computing the Fourier transform
of the non-axisymmetric density excess (Eq.~(\ref{eq:foursig})), the
$m=2$ mode is largely dominant for the bar. The angular displacement,
for fixed radial distance $R$, of the maxima of the $m=2$ mode at two
successive snapshots separated by a time $\Delta t$ yield the $m=2$
pattern speed at the distance $R$, namely $\Omega_2(R)\simeq
(\phi_2(R,t+\Delta t)-\phi_2(R,t)) /\Delta t$, where $\phi_2(R,t)$
is defined in Eq.~(\ref{eq:ampphase}) for $m=2$. We use 
$\Delta t=25 Myr$. We prefer this
method over extending the Fourier transform (say of $A_2$, $B_2$) in
the time domain, since the latter approach assumes that patterns are
characterized by constant frequencies in a relatively long time
window, while the bar's pattern speed appears to vary significantly
over rather short time windows (see below). Figure \ref{fig:omep65}
shows our calculation for the snaphsot $t=1.625$ Gyr. The curve
$\Omega_2(R)$ exhibits an approximate `plateau' at radii $2\sim 5$ kpc. 
We estimate the bar pattern speed $\Omega_{bar}$ as the mean
value of $\Omega_2$ in the interval $2.5\mbox{kpc}<R<4 \mbox{kpc}$ 
(the plateau is more clear in this range of radii). 
The vertical lines mark the positions of the Inner Linblad Resonance
(ILR), co-rotation (CR), and outer Linblad resonance (OLR) estimated
from the relations $\Omega(R_{ILR})-\Omega_{bar} = {1\over 2}\kappa(R_{ILR})$, 
$\Omega(R_{CR})-\Omega_{bar} = 0$, and $\Omega(R_{OLR})-\Omega_{bar} = 
-{1\over 2}\kappa(R_{OLR})$, where $\kappa(R)$ is the epicyclic frequency 
derived from $\Phi_0$. The plateau is well formed outside the ILR radius, 
and yields a pattern speed $\Omega_{bar}\simeq 42$ km/sec/kpc. More 
comments on the structure of the curve $\Omega_2(R)$ are made in section 
4.
\begin{figure}
\centering
\includegraphics[scale=0.16]{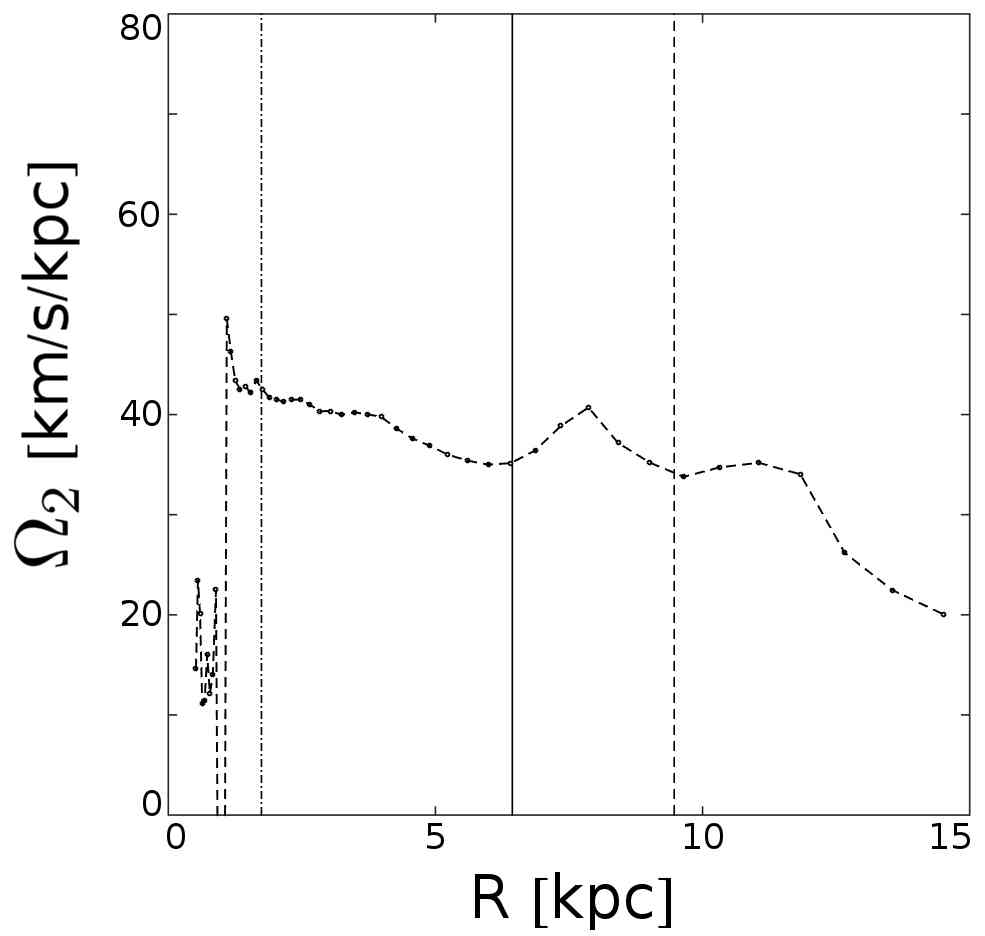}
\caption{\small The $m=2$ mode pattern speed $\Omega_2(R)$ for the snapshot 
$t=1.625$ Gyr. The bar pattern speed is estimated as the mean value of 
$\Omega_2(R)$ in the interval $2.5\mbox{kpc}\leq R\leq 4\mbox{kpc}$. The inner 
and outer dashed vertical lines mark the position of ILR and OLR. The solid 
vertical line marks the position of co-rotation (CR).} 
\label{fig:omep65}
\end{figure}

Having fixed $\Omega_{bar}$ we compute the unstable Lagrangian points
$L_1$ and $L_2$ in the Hamiltonian (\ref{eq:ham2d}). Since we use the
full N-body potential, the Lagrangian points are not found at exactly
symmetric positions with respect to the disc's center, while their
Jacobi energies have also a small difference
$2|E_{J,L1}-E_{J,L2}|/|E_{J,L1}+E_{J,L2}|\approx 10^{-4}$. Similar
differences hold for the stable Lagrangian points $L_4$ and $L_5$.  
We finally use the information of the Lagrangian points to compute the
apocentric manifold spirals emanating from the co-rotation zone.
Depending on the choice of Jacobi energy, each orbit of the PL1 or PL2
families yields a different apocentric manifold. To end up with just
one (representative) manifold calculation per family, we compute first
the orbits PL1 and PL2 for various Jacobi energies, and keep, in each
case, the orbit with Jacobi energy midway between the energies at 
$L_{1,2}$ and $L_{4,5}$, i.e., $E_{J,PL1} = (E_{J,L1}+E_{J,L4})/2$,
$E_{J,PL2} = (E_{J,L2}+E_{J,L5})/2$. These energies roughly correspond
to the median of the Jacobi energy distribution for the N-body
particles in the corotation zone. The apocentric unstable manifolds of
the corresponding orbits PL1 and PL2 are then computed, taking an
initial segment on the apocentric surface of section with N=10000
initial conditions distributed according to $\phi_j = \phi_{0,PL1} -
j\Delta\phi/N$, $p_{\phi,j} = p_{\phi,0,PL1} + j\Delta p_\phi/N$, for
the outer branch, and $\phi_j = \phi_{0,PL1} + j\Delta\phi/N$,
$p_{\phi,j} = p_{\phi,0,PL1} - j\Delta p_\phi/N$ for the inner branch
of ${\cal W}^{UA}_{PL1}$, with $\Delta\phi$ adjusted around a value
$\approx 10^{-3}$ so as to give comparable lengths for all
  computed manifolds, while $\Delta p_\phi= S_u\Delta\phi$, where
$S_u$ is the slope of the unstable eigendirection of the the monodromy
matrix at the fixed point PL1 (and similarly for the orbit PL2).

\begin{figure}
\centering
\includegraphics[scale=0.20]{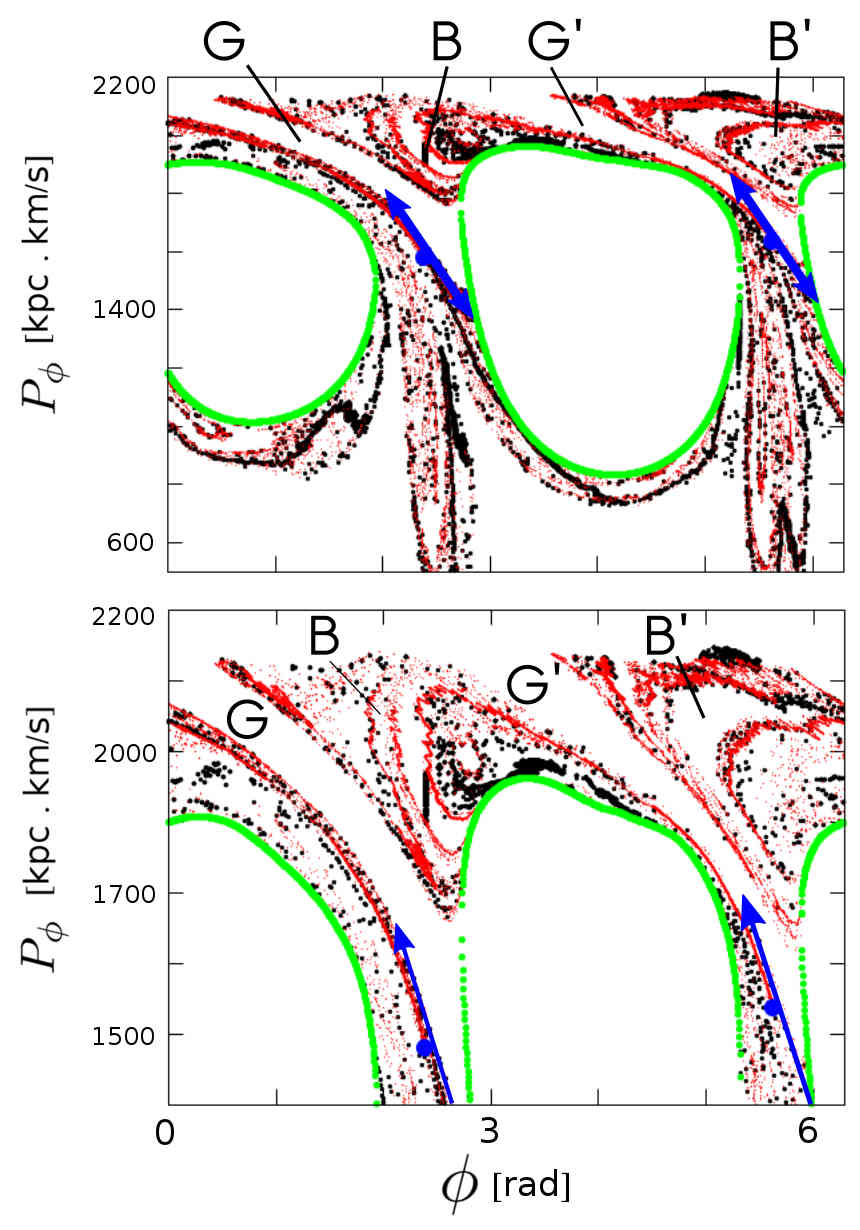}
\caption{\small The apocentric surface of section as defined in
Eq.(\ref{eq:aposos}) for the energy $E_J = 0.5(E_{J,L1}+E_{J,L4})$,
at the snapshot $t=1.625$ Gyr. The orbits in this surface of section
are strongly chaotic. The bottom panel is a magnification of
the top panel in the region of angular momenta $p_\phi$ close to the
value at corotation. The black thick points are the iterates of orbits 
with initial conditions taken at various locations within the chaotic domain. 
The red points show the the apocentric invariant manifolds 
${\cal W}^{UA}_{PL1,2}$. They emanate from the Lyapunov periodic orbits PL1 
and PL2 shown as two blue fixed points (left and right respectively). 
The thick blue arrows indicate the corresponding unstable eigendirections. 
The closed green thick curves are limiting curves inside which the orbits 
are energetically forbidden.  The marked features $B,B'$ and $G,G'$ are
the `bridges' and `gaps' (see text).}
\label{fig:surfsec}
\end{figure}
\begin{figure}
\centering
\includegraphics[scale=0.30]{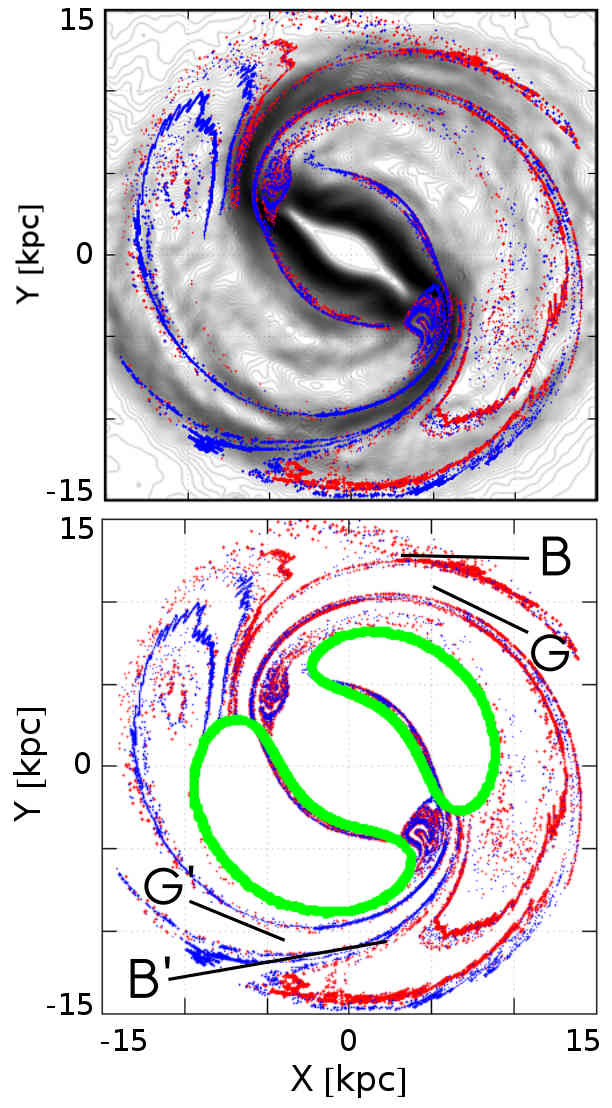}
\caption{\small Top: the apocentric manifolds ${\cal W}^{UA}_{PL1}$ (red) 
and ${\cal W}^{UA}_{PL2}$ (blue) superposed to the `Sobel-Feldman' image 
of the disc at $t=1.625$ Gyr. Besides the spirals, the manifolds reproduce 
several secondary patterns in the disc beyond the bar, such as `gaps', 
`bridges', `rings' and `bifurcations' of secondary spiral arms (see text 
for details). Bottom: the apocentric manifolds together with the limiting 
curves of the apocentric surface of section (green), and with the bridges
and gaps marked as in the corresponding surface of section of
Fig. \ref{fig:surfsec}.}
\label{fig:manall065}
\end{figure}
Figure \ref{fig:surfsec} shows the form of the apocentric surface of
section (plane $(\phi,p_\phi)$), as well as the apocentric manifolds
of the PL1 and PL2 orbits, at the snapshot $t=1.625$ Gyr. The
manifolds (red points) emanate as straight lines starting from the
fixed points corresponding to the periodic orbits PL1 and PL2, but
they soon develop a very complicated form, characterized by a number
of thin lobes forming a `chaotic tangle', which is the signature of
homoclinic dynamics. Taking segments of initial conditions in the
interior of the manifolds defined by the lobes marked $B,B'$, we
iterate these chaotic orbits (black points), and find that the
distribution of the iterates covers more and more uniformly the area
inside the manifolds' lobes.

All together, the phase portrait formed at the indicated level of
energies (Jacobi constant $E_J = 0.5(E_{J,L1}+E_{J,L4})$) has 
considerable chaos. Chaos can be partly due to numerical effects, 
(e.g. the `microchaos' generated by noise in the potential computation), 
but it is also of dynamical origin, due to the strong bar and spiral 
amplitudes. In fact, while chaos fills a substantial part of the 
phase space at corotation already in models with strong bars (see, 
for example, \cite{patetal1997}), the deformation of all phase space 
structures due to the spiral mode is evident in Fig. \ref{fig:surfsec}.  
This deformation affects the form of 
the closed limiting curves in the apocentric surface of section, inside
which the motion is energetically forbidden.  These curves are found
by locating the pairs $(\phi,p_\phi)$ for which the effective potential 
satisfies the relation
\begin{equation}
\Phi_{\rm eff}(R_c;\phi,p_{\phi})={p_\phi^2\over 2R_c^2}-\Omega_{bar}p_{\phi} 
+\Phi(R,\phi) = E_J
\end{equation}
where $R_c$ is the root of the equation $\partial\Phi_{eff}/\partial R=0$. 
The above limiting curves are similar in shape to the usual curves of zero 
velocity, but provide the most stringent limits for the apocentric surface 
of section (see \cite{tsouetal2009}. 

In Fig. \ref{fig:surfsec} the manifolds' lobes from PL1 and PL2 
encircle the corresponding limiting curves. However, despite the lack of 
energetic barrier, we observe that the uppermost manifold lobes do not 
fill the whole area around the limiting curves, but perform a number 
of oscillations leaving narrow gaps, denoted $G,G'$, in the surface of 
section. This behavior of the manifolds is dictated by basic rules of 
dynamics, which assert that the manifold lobes of the same or different 
periodic orbits cannot intersect each other. In this way, we observe that 
the lobes of the manifold emanating from PL2 approach the manifold emanating 
from PL1 in the area marked $B$, but do not intersect it, leaving instead a 
gap $G$. Similarly, the lobes of the manifold from PL1 approach the manifold 
from PL2 in the area $B'$, leaving a gap $G'$. These features have a specific 
morphological correspondence in physical space, as shown below. Let us note 
also that the level of Jacobi energy of the surface of section in 
Fig.  \ref{fig:surfsec} is close to the median of the entire distribution of 
the simulation's particles in chaotic orbits, which is consistent with the 
general expectation that most particles supporting chaotic spirals should 
be distributed in the interval of energies between $E_{J,L1}$ and $E_{J,L4}$ 
(\cite{pat2006}; \cite{tsouetal2008})).  

Figure \ref{fig:manall065} shows, now, the main result regarding the 
comparison between the disc morphology and the apocentric manifolds at the 
snapshot $t=1.625$ Gyr. The Sobel-Feldman image of the N-body disc shows a 
nearly bi-symmetric set of spirals, along with secondary ring and spiral 
features beyond the bar. The gaps $G$ and $G'$ appear as narrow zones 
separating the manifolds at the bridges $B$, $B'$. In these bridges, the 
manifold emanating from the region of $L_1$ approaches, in a nearly-tangent 
direction, the exterior side of the manifold emanating from the region of 
the Lagrangian point $L_2$, and vice versa. The approach takes place via 
several oscillations of the manifolds in space, forming patterns recognized 
in the plot as bundles of preferential directions occupied by the manifolds. 
Such bundles mostly form spiral patterns, while breaks, or `bifurcations',
are also observed, splitting in two some of these bundles. Most notably, 
the main morphological features of the apocentric manifolds have counterparts 
in the Sobel-Feldman image of the real patterns formed by the disc particles.

As a final remark in this section, the manifolds ${\cal W}^{UA}_{PL1}$
and ${\cal W}^{UA}_{PL2}$ exhibit some lopesidedness, manifest also in
the form of the limiting curves of the apocentric surface of section
as projected in physical space. This effect implies that the outflow
of particles in chaotic orbits, in the directions indicated by the
manifolds, is not symmetric with respect to the disc center. Since the
phase space outside corotation is open to escapes, particles escaping
in chaotic orbits carry with them linear momentum in a distribution of
orientations non-symmetric with respect to the center. The
contribution of this to maintaining a disc-halo `off-centering effect'
is discussed in subsection 4.3.

\section{Secular evolution}

In this section we focus on how the various spiral and other
non-axisymmetric features beyond the bar evolve along with the bar's
secular evolution.  We also examine which mechanisms are responsible
for the maintainance of appreciable levels of non-axisymmetric
activity. In the end of the section we return to the comparison
between manifolds and the observed patterns in the disc (as in
Fig.~\ref{fig:manall065}), but for different snapshots covering the
whole disc's secular evolution.

\subsection{Bar spin-down}
\begin{figure}
\centering
\includegraphics[scale=0.12]{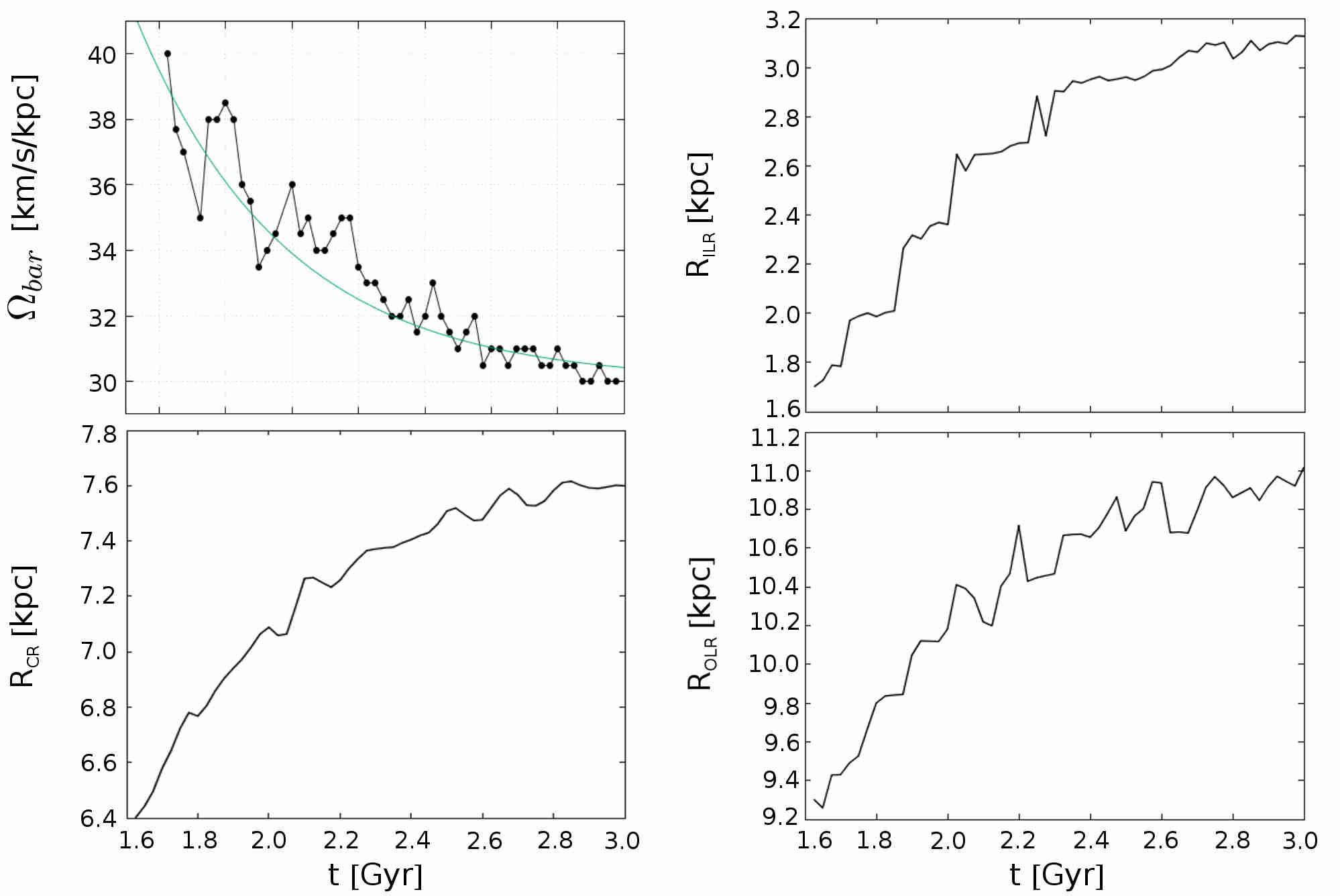}
\caption{\small Top left: Evolution of the bar's pattern speed. The
cyan curve represents an exponential fitting $\Omega_{bar}(t)
=\Omega_0-\Omega_1\left(1-\exp\left(-{t-t_0\over t_d}\right)\right)$, 
with $\Omega_0=42$ km/sec/kpc, $\Omega_1=11$ km/sec/kpc, $t_0=1.625$ Gyr, 
$t_d=2$ Gyr.  The remaining panels show the evolution of the radii of the 
bar's ILR (top right), CR (down left), and OLR (down right).}
\label{fig:omegadecay}
\end{figure}

A basic manifestation of secular evolution is the bar's spin down, as
shown in Fig. \ref{fig:omegadecay}, which leads to a an outward slow
displacement of the ILR, CR and OLR. Thus, $(R_{ILR},R_{CR},R_{OLR})
\simeq (1.75,6.5,9.5)$ kpc immediately after the bar formation and 
$(3.1,7.5,12)$ kpc at the end of the simulation. The bar's spin down 
can be associated with several factors, as for example: i) particles 
abandoning the bar and carrying angular momentum outwards at incidences 
of growing spirals (see below), or ii) transfer of angular momentum to 
the halo (see \cite{debsel1998}; \cite{debsel2000}; \cite{ath2002};
\cite{athmis2002}; \cite{ath2003}). 

\begin{figure}
\centering
\includegraphics[scale=0.18]{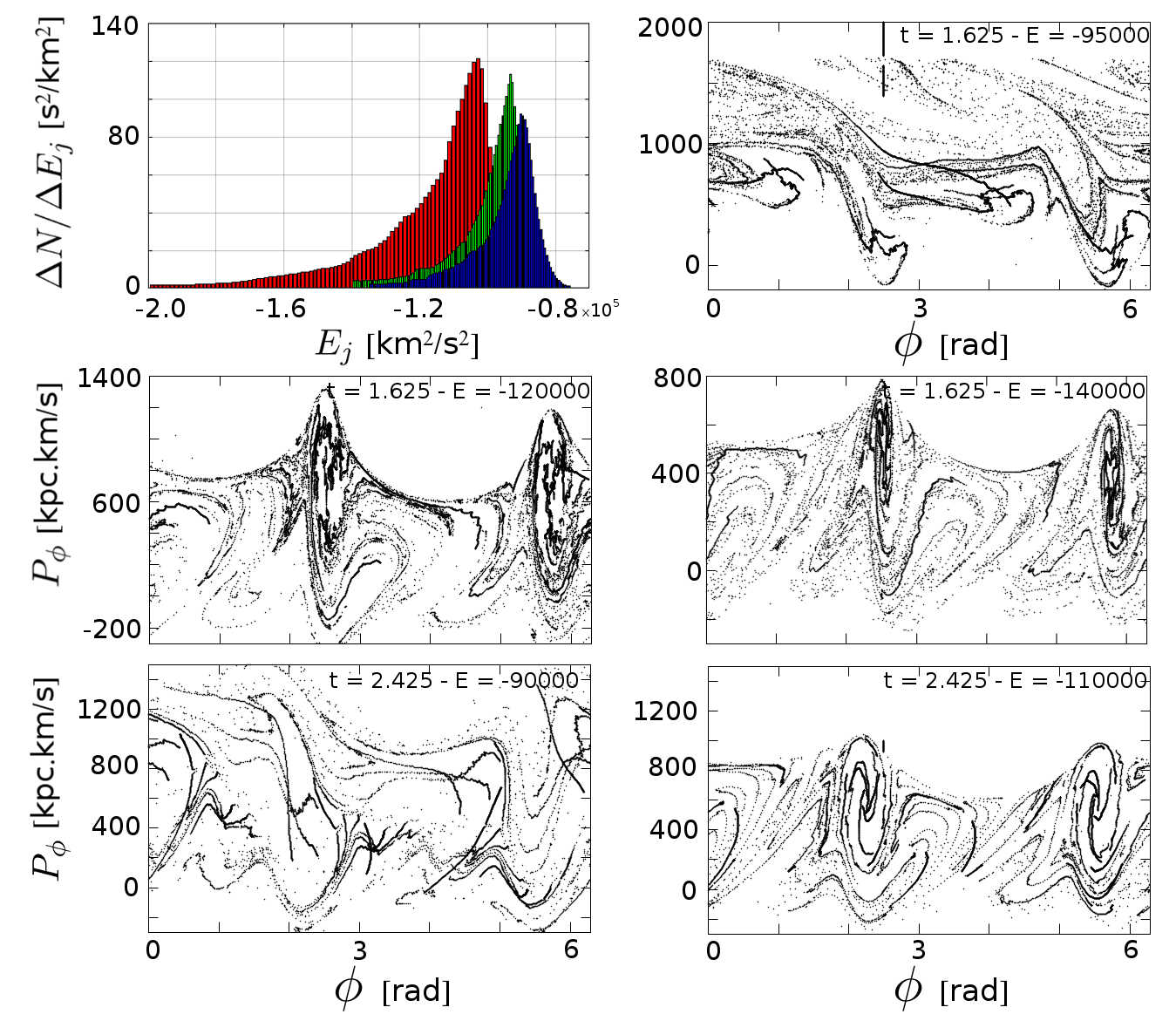}
\caption{\small Top left: Evolution of the distribution of the Jacobi energies 
of the disc particles at the snapshots $t=1.625$ Gyr (red), $2.05$ Gyr (green), 
and $2.425$ Gyr (blue). In comparison, the jacobi energies at CR are 
(in $km^2/s^2$), $E_J=-1.05\times 10^5$, $-0.94\times 10^5$ and 
$-0.91\times 10^5$ respectively.  The top right and middle panels show the 
apocentric surfaces of section for the snapshot $t=1.65$ Gyr, in energies 
different from the one of Fig.~\ref{fig:surfsec}, selected to cover the range 
of the corresponding distribution. The bottom panel shows two surfaces of
section for the snapshot $t=2.425$. Chaos is ubiquitous in all these plots.}
\label{fig:chaos}
\end{figure}

The bar retains high amplitudes throughout the simulation. Thus, whereas 
slowing down, the bar sustains appreciable levels 
of orbital chaos in the disc. Figure \ref{fig:chaos}, top panel, shows the 
evolution of the distribution of the disc particles' Jacobi energies, along 
with more phase portraits at energies, or snapshots, different from the one 
of Fig.~\ref{fig:surfsec}. The distribution of Jacobi energies evolves
by keeping its global maximum always close to the instantaneous CR value, 
hence it is shifted towards absolutely smaller energies as the bar slows down. 
The dispersion of the Jacobi energies is also reduced in time, but remains at 
a level of few times $10^4km^2/s^2$. The remaining panels in 
Fig.~\ref{fig:chaos} show apocentric surfaces of section selected so as to 
cover the entire range of energies where particles are practically distributed, 
for $t=1.65$ Gyr (as in Fig.~\ref{fig:surfsec}) and at a later snapshot
$t=2.425$ Gyr. Similar phase portraits are found in all snapshots. Chaos is 
ubiquitous in all these portraits, indicating that the spiral and other 
structures beyond the bar are supported mostly by chaotic orbits.

\subsection{Incidents of spiral and other non-axisymmetric activity}

\begin{figure}
\centering
\includegraphics[scale=0.2]{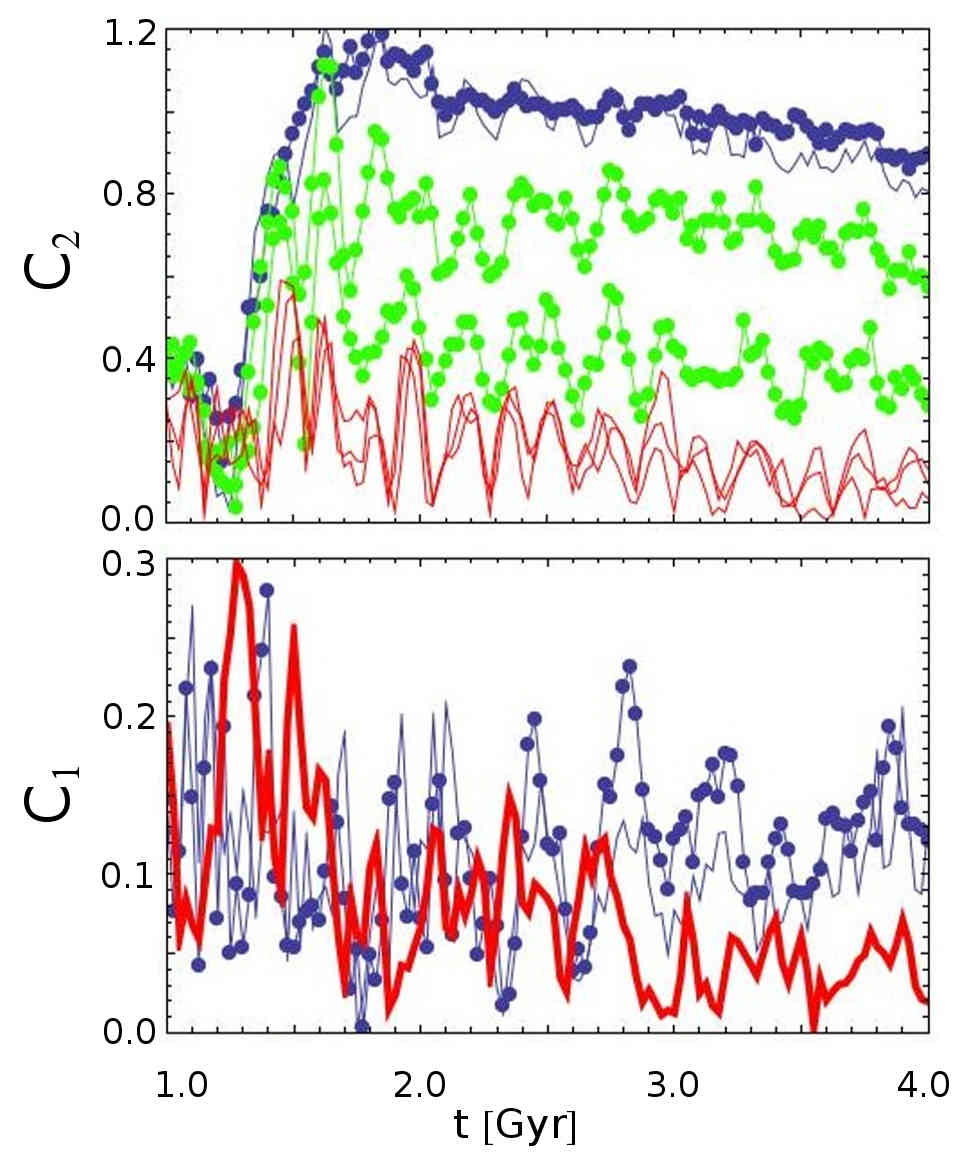}
\caption{\small Top: evolution of the Fourier amplitude $C_2$ in the interval 
$1\leq t[Gyr]\leq 4$. The value of $C_2$ is shown at the distances $R=3,4$ kpc 
(dotted blue), $R=5$ kpc (solid line, blue), $R=6,7$ kpc (dotted green), and 
$R=8,9,10$ kpc (solid red). Bottom: evolution of the Fourier amplitude $C_1$ 
at $R=3$ kpc (dotted blue), $R=4$ kpc, (solid blue), and $R=12$ kpc (thick 
solid red).} 
\label{fig:m12evolve}
\end{figure}

The variability of the bar, spiral and other non-axisymmetric patterns is 
reflected in time variations of the Fourier amplitude $C_2(R)$, as shown 
in Figure \ref{fig:m12evolve}, top panel, for different radii. The group 
of curves for $R=3,4$ and $5$ kpc quantify the evolution of the bar's 
strength.  The bar reaches its maximum strength near $t=1.6$ Gyr, leading 
to a $m=2$ amplitude larger than unity. Thereafter, the bar $m=2$ amplitude 
stabilizes in the interval from $t=1.6$ to $t=2.1$ Gyr. After a small but 
abrupt drop at $t=2.1$, associated with the buckling instability, the bar's
amplitude slowly decays through the remaining $~2$ Gyr of the simulation, 
remaining however always of order of unity.

The increase and subsequent slow decay of the $m=2$ bar mode permeates
all radial annuli across the disc. The $m=2$ amplitude in general
falls with $R$ between CR and OLR, but it tends to stabilize to a mean
level $\sim 0.2$ in an annulus $8\leq R[kpc]\leq R_{OLR}$. The
amplitude $C_2$ undergoes oscillations which are nearly in-phase for
all radii within this annulus.  At local maximum of these oscillations
$C_2$ exceeds 0.2, up to $t\sim 3$ Gyr, while the oscillations are
milder afterwards. At all peaks of the $C_2$ mode in the interval
$1.5\leq t[Gyr]\leq 4$ we observe `incidents' of non-axisymmetric
activity in the outer parts of the disc. Taking the mean $C_2$ from the 
three lowermost curves in the top panel of Fig.~\ref{fig:m12evolve}, 
we estimate the times when such incidents reach maximum amplitude 
(to a precision related to the frequency of saving of the particles' 
positions, i.e., every 25 Myr). 
We thus identify the following sequence of times:
\begin{eqnarray}\label{eq:tseqburst}
t[Gyr]^{(incident)} = 1.625, 1.8, 1.95, 2.2, 2.375, 2.525,\nonumber\\
2.725, 2.950,3.1, 3.325,3.75, 3.95.
\end{eqnarray}
By the above sequence, an approximate period $T\approx0.2$ Gyr can be
deduced, but with large fluctuations $\Delta T\sim 0.1$ Gyr. The bar has 
a pattern speed $\Omega_{bar}\approx 40$ Km/sec/kpc (period $T_{bar}\approx 
0.15$ Gyr) at the starting time and $\Omega_{bar}\approx 30$ Km/sec/kpc 
($T_{bar}\approx 0.2$ Gyr) at the end of the sequence. Thus, the sequence 
(\ref{eq:tseqburst}) is in rough resonance with the bar's rotation. However, 
the bar itself exhibits some extra phenomena. The bottom panel of
Fig. \ref{fig:m12evolve} show important oscillations in the $m=1$ mode at 
disc radii smaller than the bar's half-length.  In these oscillations $C_1$ 
reaches a mean level $\sim 0.1$, with peaks up to $\sim 0.3$, and an 
approximate period $T\sim 0.4$ Gyr in the interval from $t=2$ Gyr to 
$t=3$ Gyr. Oscillations of $C_1$ appear also in the outer disc, albeit 
not in phase with those of the inner disc.

\begin{figure}
\centering
\includegraphics[scale=0.17]{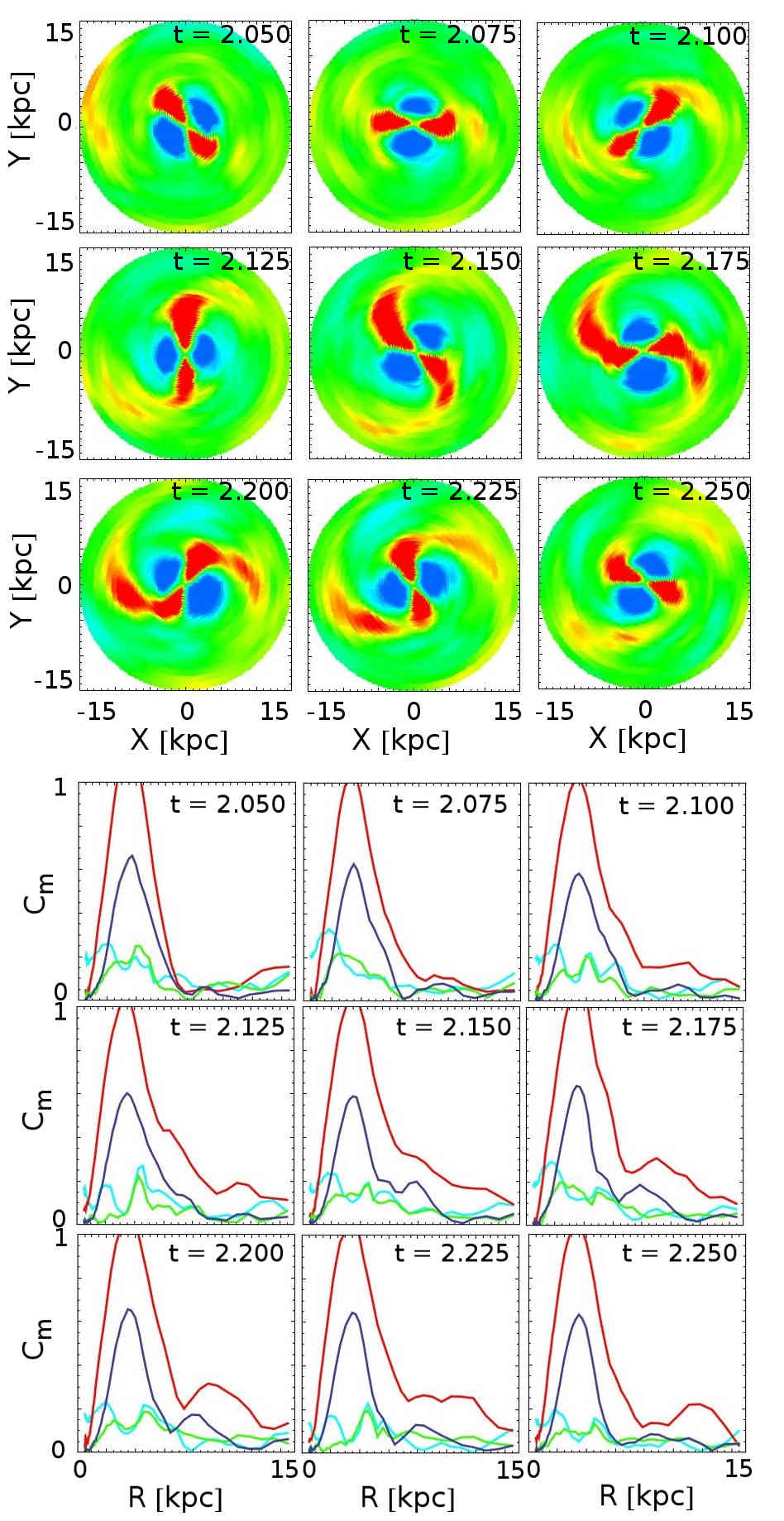}
\caption{\small Top: an `incident of inner origin', whose corresponding 
$C_2$ maximum occurs around $t=2.2$. Bottom: the Fourier amplitudes $C_m(R)$ 
for $m=1$ (cyan), $m=2$ (red), $m=3$ (green), and $m=4$ (blue). The dotted 
vertical line marks the position of the bar's CR. An $m=2$ disturbance 
propagates from CR outwards. Significant rise of the $m=1$ and $m=3$ modes 
appears inside the bar at the initial phase of the incident.} 
\label{fig:inburst}
\end{figure}

\begin{figure}
\centering
\includegraphics[scale=0.17]{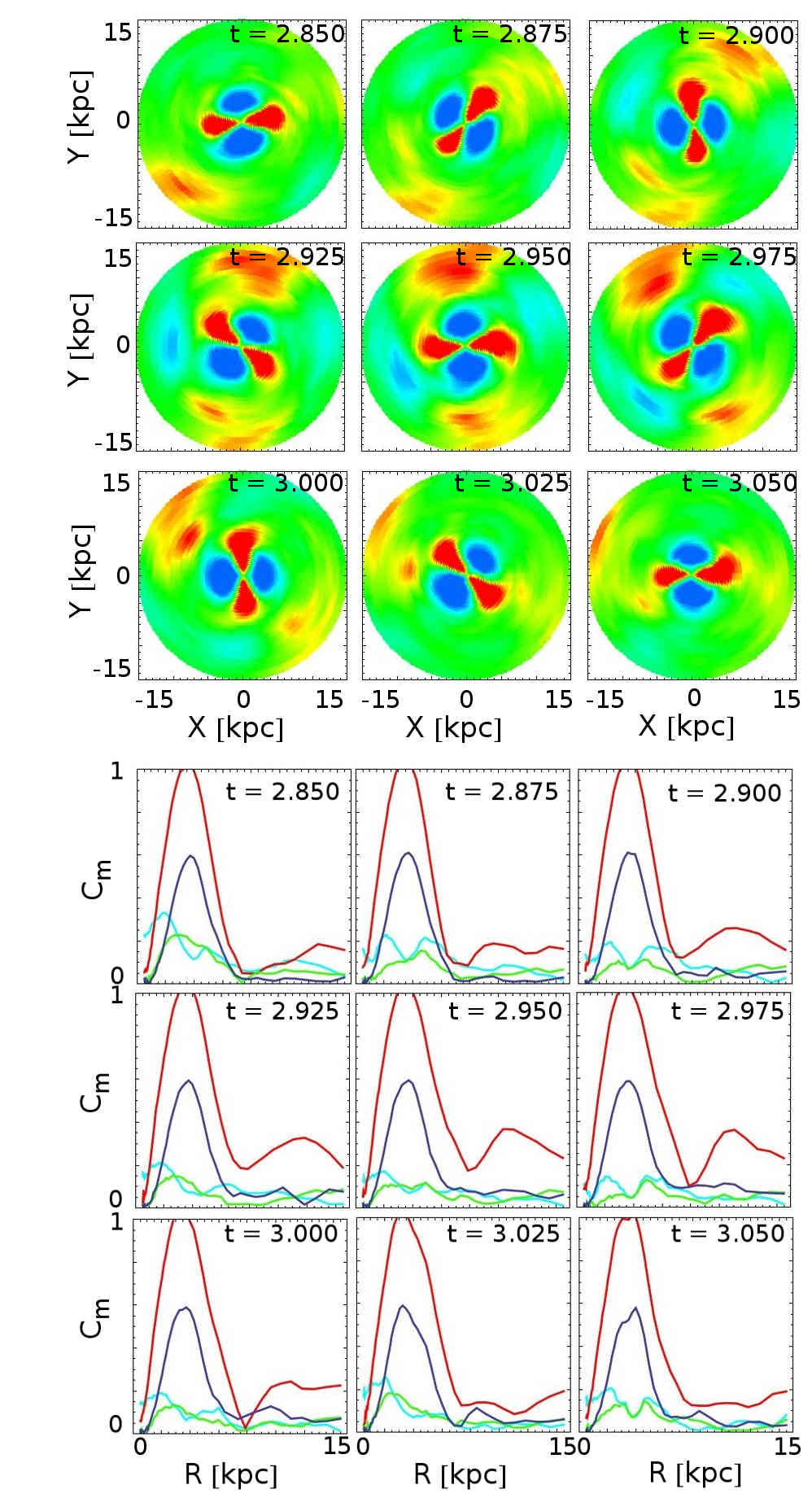}
\caption{\small Same as in Fig.~\ref{fig:inburst}, but for an incident of 
outer origin reaching its peak around $t=2.95$. An $m=2$ disturbance 
generated in the outer parts of the disc is reflected at CR. Significant 
rise of the $m=1$ and $m=3$ modes appears again inside the bar.} 
\label{fig:outburst}
\end{figure}

Through density excess maps, and the corresponding profiles of the Fourier 
modes $m=1,2,3,4$, we can see how incidents of non-axisymmetric activity 
affect the disc's morphology beyond the bar. We distinguish two types of 
incidents, exemplified in Figs.\ref{fig:inburst} and \ref{fig:outburst} 
respectively, as follows:

-{\it Incidents of inner origin:} an incident of inner origin
corresponds morphologically to a spiral wave emanating from the end of
the bar, and travelling outwards, until it becomes detached from the
bar. An example is given in Fig. \ref{fig:inburst}, showing the
density excess maps at nine snapshots around the time $t=2.2$, which
belongs to the sequence (\ref{eq:tseqburst}). The outwards `emission'
of the spiral wave from the bar becomes evident after $t=2.15$ Gyr. At
earlier times, one observes the appearance of small {\it leading}
extensions of the bar, so that the whole incident is reminiscent of
swing amplification. The creation of a leading component can be
associated with flow of material approaching the bar from the leading 
direction by moving along outer families of periodic orbits such as the 
2:1 family (Patsis; private communication). The spirals start fading out 
after detaching from the bar.

- {\it Incidents of outer origin:} an incident of outer origin corresponds 
morphologically to a rise of the surface density excess in the outer parts 
of the disc, as shown in Fig. \ref{fig:outburst}. The rise appears 
both in the $m=1$ and $m=2$ modes. The left panels in Fig. \ref{fig:outburst} 
indicate that the $m=2$ perturbation arises at radial distances beyond $R=12$ 
kpc, it travels inwards,  and then  it is reflected at corotation. It is 
stressed that these features are not connected with the present analysis 
of spirals based on manifolds. However, they may contribute to other effects 
observed in the disc as discussed below.

\begin{figure}
\centering
\includegraphics[scale=0.18]{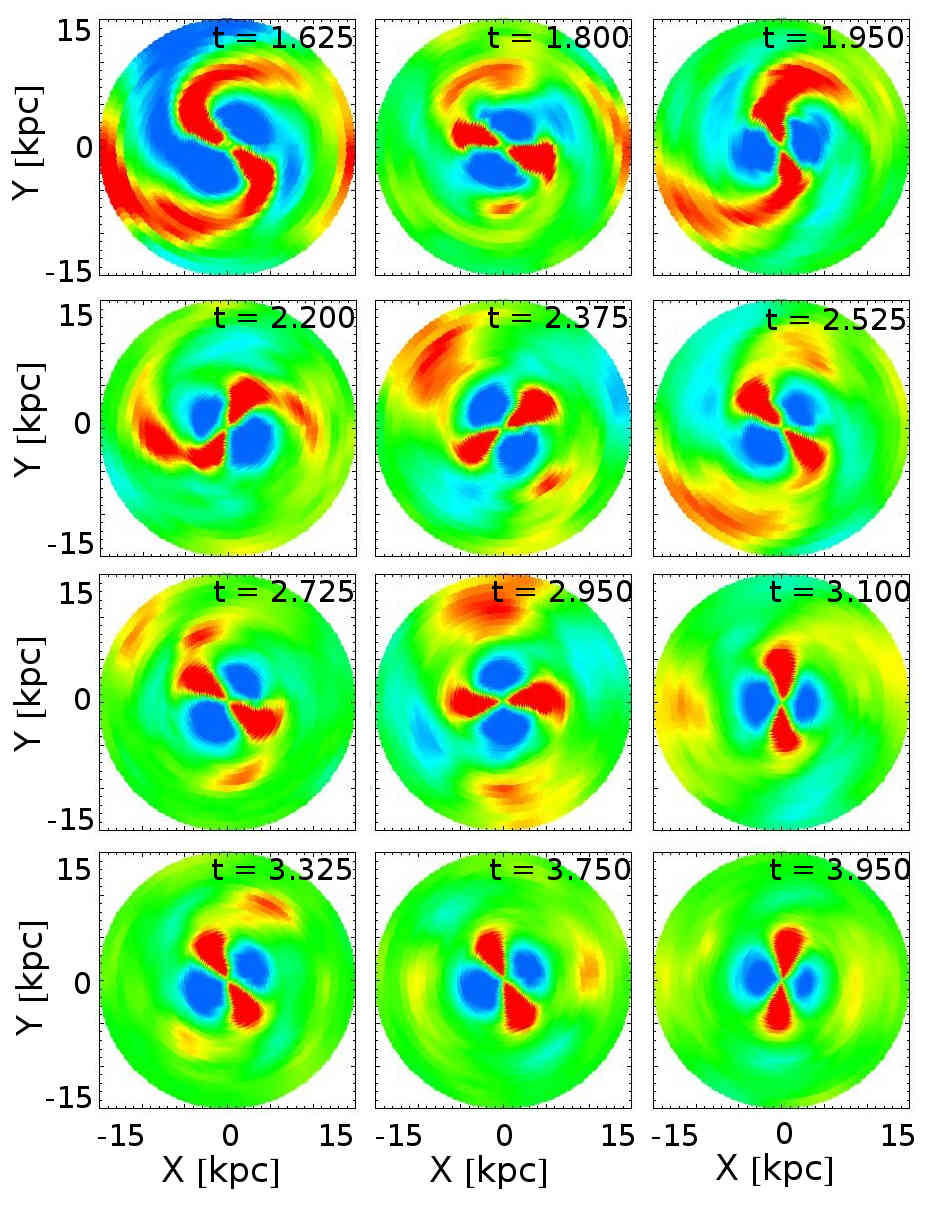}
\caption{\small Non-axisymmetric density excess maps showing the peaks of 
the `incidents' of non-axisymmetric activity for all snapshots of the sequence 
of Eq.~(\ref{eq:tseqburst}).}
\label{fig:densexmax}
\end{figure}

Figure \ref{fig:densexmax} shows the density excess map of the disc
for all times of the sequence (\ref{eq:tseqburst}). Incidents of inner
origin appear at the snapshots $t=1.625$, $t=1.95$, $t=2.2$, $t=2.725$ Gyr, 
while incidents of outer origin appear at $t=2.375$, $t=2.95$, $t=3.75$, 
and $t=3.95$ Gyr.  A classification as inner or outer is unclear for the 
incidents taking place at $t=1.8$, $t=2.525$, $t=3.1$, and $3.325$ Gyr. 
Incidents of inner origin appear mostly in the interval $1.6<t[Gyr]<3$, 
while incidents of outer origin are observed throughout the simulation. 
In fact, the disc becomes overall less responsive to non-axisymmetric
perturbations as the time goes on, and a transition from stronger to
weaker incidents is discernible around $t\approx 3$ Gyr.

\subsection{Relative disc-halo orbit and its effect on the disc}
\begin{figure}
\centering
\includegraphics[scale=0.11]{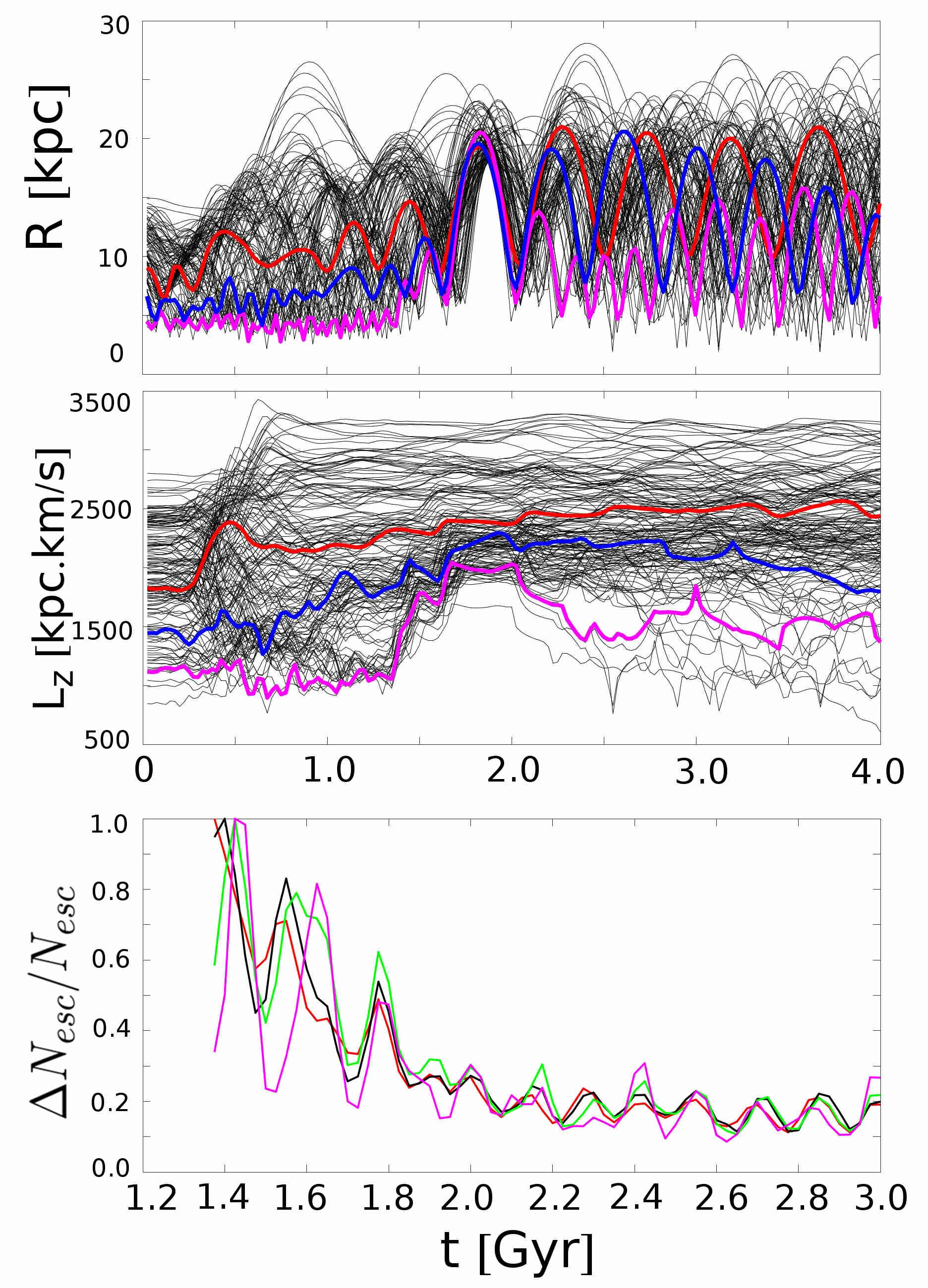}
\caption{\small Top: evolution of the radius $R$ for one hundred randomly 
selected particles whose orbits cross the annulus $R_{min}\leq R\leq R_{max}$ 
in at least one time window $W(t)$ (see text). The colored curves show three 
different types of behavior, namely, a gradual increase of the apocenter 
(red), an increase taking place mostly in isolated `incidents' (blue), and an 
orbit going to a larger and then returning to a smaller apocenter 
(purple). Middle: evolution of the angular momentum per unit mass $L_z=p_{\phi}$ 
for the same particles. Bottom: the relative fractions 
$\Delta N_{esc}(t)/N_{esc}$ (see text) when $R_{max}$ is taken equal to $12$ kpc 
(red), $14$ kpc (black), $16$ kpc (green), $18$ kpc (purple). } 
\label{fig:escall}
\end{figure}

At every inner incident of spiral activity, the particles's orbits change 
as the potential develops rapid fluctuations. This is accompanied by changes 
in the particles' energies and angular momenta. Figure \ref{fig:escall} 
summarizes these effects. We consider moving time windows 
$W(t)=(t-\Delta T_w/3,t+2\Delta T_w/3)$ with $\Delta T_w=0.125$ Gyr, as well 
as an annulus $R_{min}\leq R\leq R_{max}$, with $R_{min}=5$kpc and 
$R_{max}=18$ kpc, equal to six disc exponential scale lengths. At a snapshot 
$t$ we isolate all particles whose orbits have crossed both limits of the annulus 
within the time window $W(t)$. We finally collect these particles for all time 
windows, and check their orbits. This process collects about 10\% of the 
disc particles whose orbits behave as above at least in one time window. 
Figure \ref{fig:escall}, top, shows $R(t)$ for a sample of 100 randomly 
chosen trajectories in this set. Analogous pictures are obtained for any 
other random choice. Note that the N-body code represents forces sufficiently 
smoothly to produce smooth particle orbits. The main feature of these orbits 
is their change of elongation (difference between apocenter and pericenter) 
in different time windows. Abrupt changes are observed for some orbits at 
particular times, connected to an abrupt change in the angular momentum 
(middle panel). These times correspond to the growth phase of major incidents 
of non-axisymmetric activity in the disc. Τhe effect is more notorious for 
the first incident of spiral growth a little before $t=0.5$ Gyr (before the 
bar), as well as the growth of the manifold spirals at $t=1.5$ and $t=1.65$ 
Gyr. In every incident the majority of affected orbits become more elongated 
after the incident than before (although both types or orbital changes are 
observed). Also, the majority of affected orbits exhibit increase of the 
angular momentum, which mostly affects an orbit by increasing its apocentric 
distance to a large value (more than 20 kpc for some particles in
Fig.~\ref{fig:escall}), while many particles move also to orbits with 
pericentric distance beyond co-rotation.

The relative influence of each incident in the orbits can be estimated 
by measuring the fraction $\Delta N_{esc}(t)/N_{esc}$, where  $N_{esc}$ 
is the total number of particles which, in at least one time window exhibited 
a crossing of the annulus $R_{min}\leq R\leq R_{max}$ as described above, 
and $\Delta N_{esc}(t)$ is the number of particles doing so in the time 
window $W(t)$. The bottom panel in Fig.\ref{fig:escall} shows the evolution 
of $\Delta N_{esc}(t)/N_{esc}$ for four choices of $R_{max}$, namely 
$R_{max}=12,14,16$ and $18$ kpc. Consecutive `incidents' of non-axisymmetric 
activity cause an oscillatory behavior of these curves. For nearly all 
particles in the population $N_{esc}$ the first event is associated with the 
growth of the bar (at $t=1.4$ Gyr), while for most particles (up to a percentage 
$80$\%) we have repeated events associated with the growth of the manifold 
spirals at $t=1.5$, $t=1.65$ and $t=1.8$ Gyr. Smaller percentages 
appear at subsequent times, as the incidents of non-axisymmetric activity 
beyond the bar slowly decay in amplitude.

\begin{figure}
\centering
\includegraphics[scale=0.2]{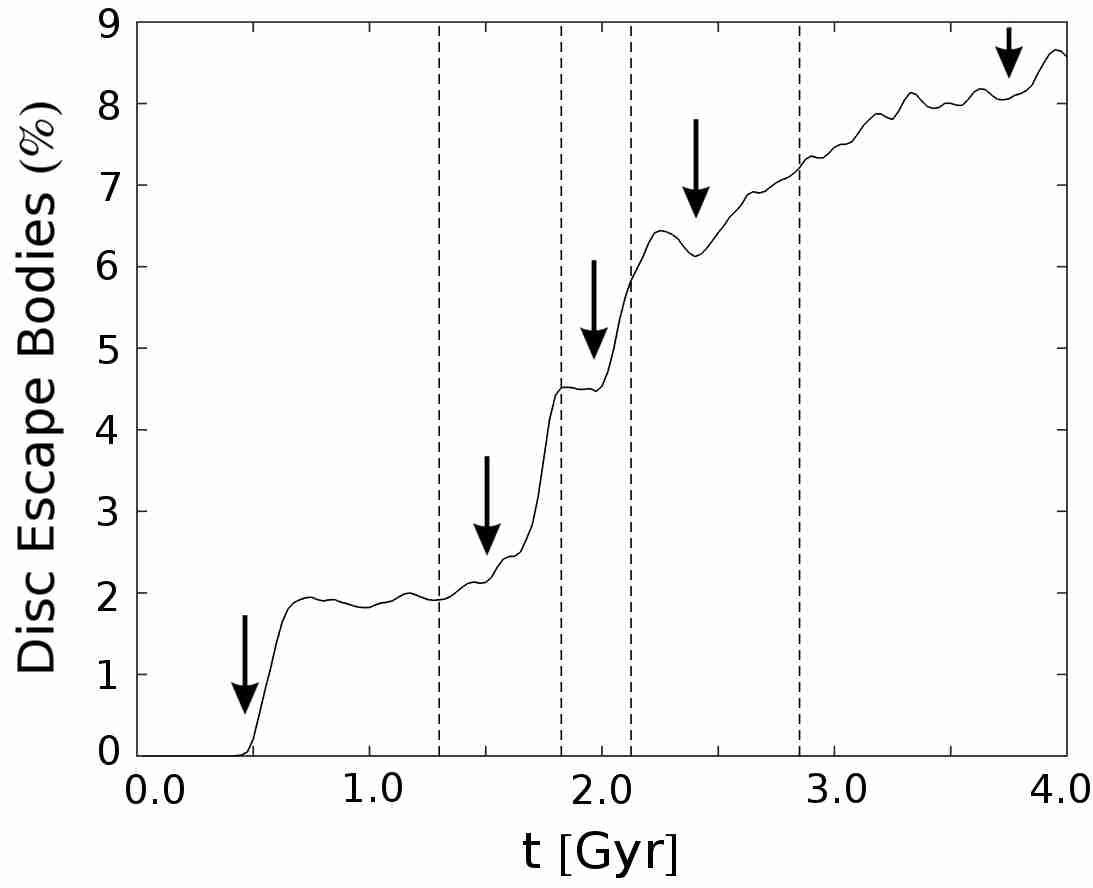}
\caption{\small Percentage of disc particles found at a distance 
$R>R_{esc}=18$ kpc, as a function of time. The arrows indicate times when 
an abrupt increase of escaping disc particles is observed. The dashed 
vertical lines correspond to the limits of time intervals ($t_1$ to $t_4$) 
in which the relative disc-halo orbit is observed to change character, i.e., 
to switch from circumcentric to epicyclic, and vice versa (see text). } 
\label{fig:escdisc}
\end{figure}
The orbital changes observed at major incidents of non-axisymmetric
activity result in a increase in time of the percentage of disc particles 
beyond any sufficiently large radius $R$. The increase is statistical, 
since it is due both to particles really escaping the system or ones moving 
in bounded, but elongated orbits, being pushed, however, outside co-rotation. 
We conventionally call the radius $R=R_{esc}=18$ kpc as the `escape' radius, 
and the particles found at $R>R_{esc}$ at the time $t$ `escape bodies'. 
Figure \ref{fig:escdisc} shows the evolution of the percentage of `escape
bodies' $\Delta N_{esc}/N_{disc}$.  The contribution of major incidents 
of non-axisymmetric activity is identified in this plot as an abrupt increase 
of $\Delta N_{esc}/N_{disc}$. Three major events are observed to take place 
at the times $t=0.55$, $t=1.7$ and $t=2.1$ Gyr, while smaller ones at the 
times $t=2.5$ and $t=2.9$ Gyr. Besides the first event at $t=0.55$ (before 
the bar), the remaining ones are all associated with incidents of inner 
origin. Outer incidents may also affect the particles' orbits, but their 
influence on $\Delta N_{esc}/N_{disc}$ should be smaller, as the
corresponding disc disturbances propagate inwards and then outwards
(being reflected at CR, see above).

\begin{figure}
\centering
\includegraphics[scale=0.22]{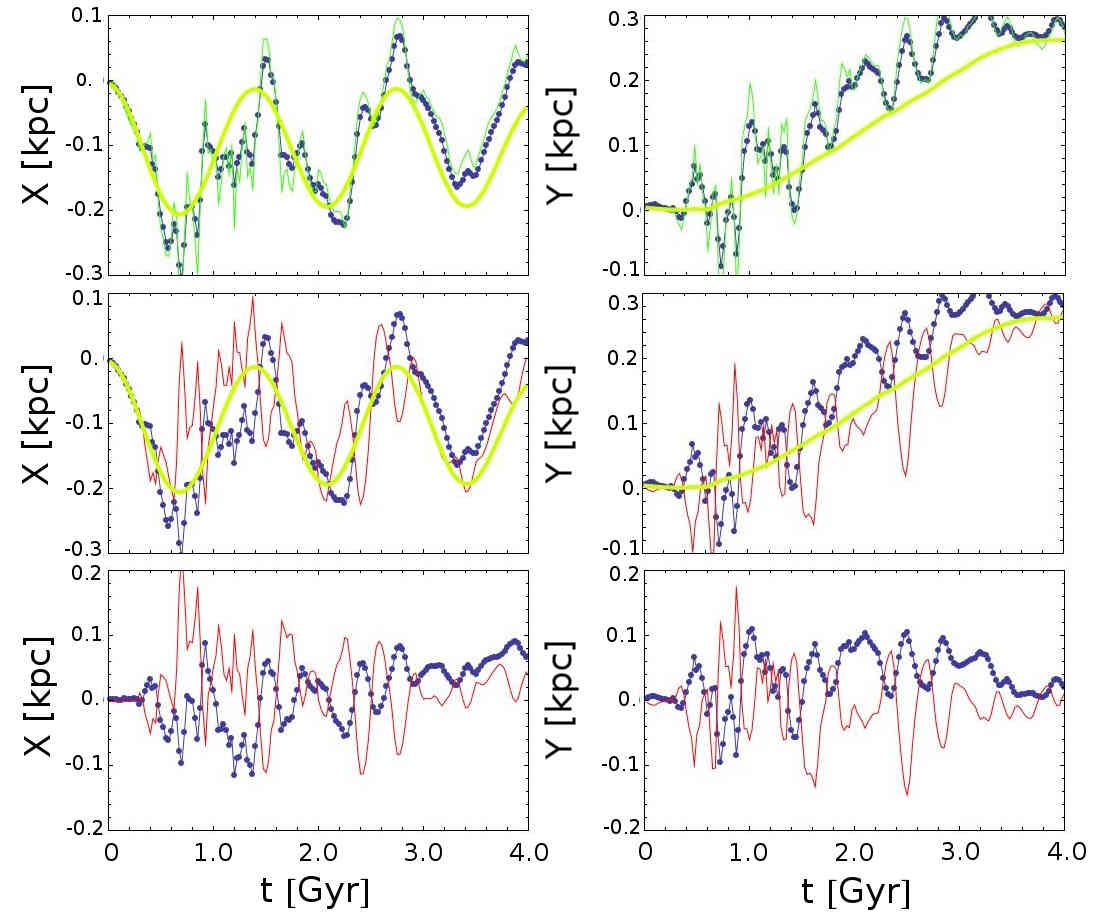}
\caption{\small Top row: Time evolution of the $x-$coordinate (left), and 
$y-$coordinate (right) of the center of mass of the halo particles (blue 
dotted) or the bulge particles (cyan solid line). The thick green line 
shows the corresponding evolution of the $x-$ and $y-$ co-ordinates of 
the total center of mass of all particles included in the simulation box. 
Middle row: same as in the top row but for the centers of mass of the 
halo particles (blue-dotted) and the disc particles (red solid). Bottom 
row: Relative orbits of the halo (blue dotted) and disc (red solid) 
centers of mass with respect to the center of mass of the total system.} 
\label{fig:habgdscxy}
\end{figure}
Assuming, now, that the particles connected with these events move in 
chaotic orbits, one expects important fluctuations in the number of particles 
escaping co-rotation as a function of the final {\it direction} in which the 
particles move. As a result, the main body of the disc {\it recoils} in a 
direction changing stochastically in time, the effect being larger at times
connected with major incidents. This recoil sets the main body of the disc 
in relative orbit with respect to the spheroid components (i.e. bulge+halo).  
Figure \ref{fig:habgdscxy} illustrates the phenomenon. We compute the evolution 
of the position of the center of mass $\mathbf{r}_{cm}^{(d)}(t)$, 
$\mathbf{r}_{cm}^{(b)}(t)$, $\mathbf{r}_{cm}^{(h)}(t)$, for the disc, 
bulge and halo components respectively. For the disc particles we use an 
iterative algorithm to specify the center of mass of only those particles 
characterized as `non-escaping'. Namely, we first compute the center of mass
$\mathbf{r}_{cm}^{(d,0)}$ of all disc particles, and then consider the
center of mass of only those disc particles satisfying $R<R_{esc}$,
where $R$ is the cylindrical radial distance of a particle with respect to 
$\mathbf{r}_{cm}^{(d,0)}$. Alternative ways to compute $\mathbf{r}_{cm}^{(d)}$, 
taking the center of the computational box or the center of mass of the total 
system as origin, yield practically indistinguishable results. Finally, 
we compute the time evolution of the center of mass of all bodies in the 
simulation $\mathbf{r}_{cm}^{(all)}$. In a perfect momentum-preserving 
scheme, the vector $\mathbf{r}_{cm}^{(all)}$ should remain constant in time
(the net total momentum in the simulation initially vanishes). However, 
very slow smooth variations of the global center of mass are obtained as 
a result of the use of adaptive mesh-refinement as well as the fact that 
the accelerations of particles escaping the computational box are computed 
with multipole formulas instead of cloud-in-cell interpolation (see
\cite{kyzetal2017a} for details). These phenomena are a priori predictable, 
and they cause the global center of mass to undergo a {\it smooth} slow 
displacement with time-varying velocity not exceeding the value $0.3$ Km/sec 
in any of its three Cartesian components during the whole simulation 
(thick solid curves in top and middle panels of Fig. \ref{fig:habgdscxy}).  
On the other hand, the main effect to be observed in the same figure is that 
each of the three {\it individual} components $\mathbf{r}_{cm}^{(d)}(t)$,
$\mathbf{r}_{cm}^{(b)}(t)$, $\mathbf{r}_{cm}^{(h)}(t)$ undergo rapid
stochastic variations around the global center of mass. We checked
that such variations are important only in the projections of the
three vectors on the disc plane, and they can be described as follows:
The $x-$ and $y-$ components of the centers of mass of the halo and
bulge undergo rapid stochastic variations around the smooth curves
$x^{(all)}(t)$, $y^{(all)}(t)$, remaining, nevertheless always closely
attached to each other. On the contrary, the $x-$ and $y-$ components
of the vector of the disc center of mass exhibit rapid fluctuations
which set the disc in relative orbit with the (nearly coinciding)
centers of mass of the halo-bulge.

\begin{figure}
\centering
\includegraphics[scale=0.1]{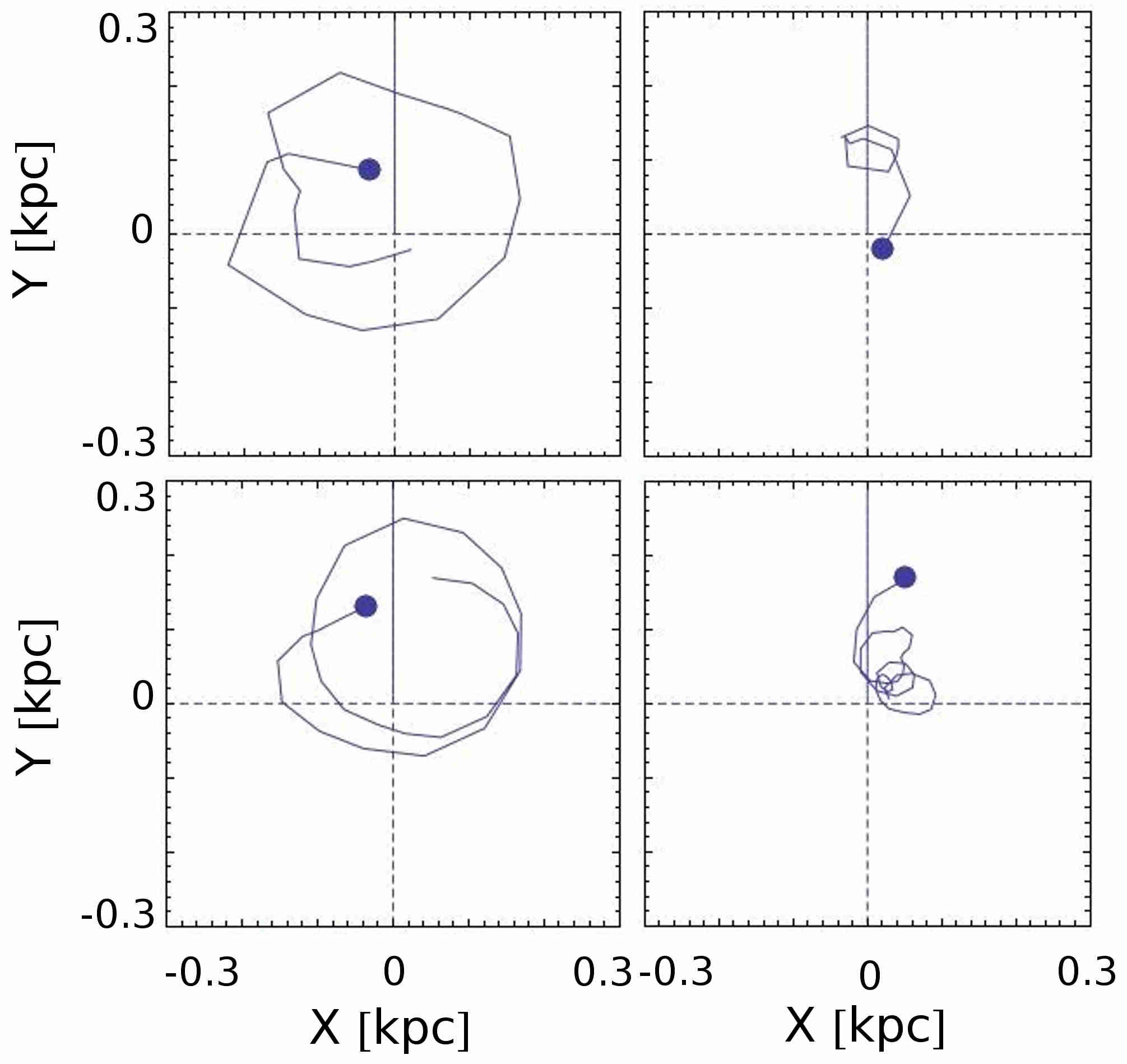}
\caption{The relative orbit of the spheroid (halo-bulge) center of mass 
with respect to a non-inertial observer centered at the disc center of mass 
(components $y$ vs. $x$ of the vector $\mathbf{\Delta r}_{hd}(t)$) at four 
time intervals $(t_1,t_2)$ (top left), $(t_2,t_3)$ (top right), $(t_3,t_4)$ 
(bottom left), and $(t_4,4\mbox{Gyr})$ (bottom right), with $t_1=1.4$ Gyr, 
$t_2=1.825$ Gyr, $t_3=2.125$ Gyr, $t_4=2.95$ Gyr.} 
\label{fig:hadscorb}
\end{figure}

The evolution of the relative orbit of the halo-bulge, and disc with
respect to the global center of mass $\mathbf{r}_{cm}^{(all)}$ is shown 
in the bottom panel of Fig.  \ref{fig:habgdscxy} ($x,y$ components of 
$\mathbf{r}_{cm}^{(h)}(t)-\mathbf{r}_{cm}^{all}(t)$ and
$\mathbf{r}_{cm}^{(d)}(t)-\mathbf{r}_{cm}^{all}(t)$). The separation between 
disc and halo-bulge increases near major incidents of non-axisymmetric activity, 
and the group of spheroidal components (halo-bulge) is set in relative motion 
with respect to the disc. The first major separation takes place near 
$t\approx 0.3$ Gyr (first major growth of spirals before the bar). 
After the bar is formed, it is unclear whether 
the bar's center of mass moves closer to the rest of the disc or to the 
spheroid (halo-bulge). We keep the conventional separation of the system 
into disc and spheroid (halo-bulge) components, hereafter referred to as 
the `relative disc-halo' orbit
$\mathbf{\Delta r}_{hd}(t)\equiv
\mathbf{r}_{cm}^{(h)}(t)-\mathbf{r}_{cm}^{(d)}(t)$.
The disc-halo orbit is depicted in Fig. \ref{fig:hadscorb}, starting from 
the time $t=1.4$ Gyr, immediately after the first incident of spiral activity 
accompanying bar formation. The four panels correspond to intervals in which 
the relative disc-halo orbit changes character, i.e., from `circumcentric' 
(described around the disc center) the orbit switches to `epicyclic' (the 
halo-bulge center of mass describes one or more small epicycles with respect 
to the disc center), and vice versa. These switches take place at the times
$t_1=1.4$ Gyr, $t_2=1.825$ Gyr, $t_3=2.125$ Gyr, and $t_4=2.95$ Gyr. The 
first `epicyclic' interval is rather short ($t_3-t_2=0.325$ Gyr), while 
the second is longer ($t_4-t_3=1.05$ Gyr) and within it the relative disc 
halo separation tends to zero, leading the whole system again close to a 
state of common center of mass of all components. Most notably, the times 
$t_1$, $t_2$, $t_3$ and $t_4$ differ by $~0.2-0.3$ Gyr, from the times when 
abrupt changes in $\Delta N_{esc}/N_{esc}$ take place in Fig.~\ref{fig:escdisc}. 
The time difference is of the order of one radial orbital period for the
particles of Fig.~\ref{fig:escall}. We regard these facts as evidence
that the orbital switches observed in Fig.~\ref{fig:hadscorb} are
connected with the disc recoil effect.

\begin{figure}
\centering
\includegraphics[scale=0.18]{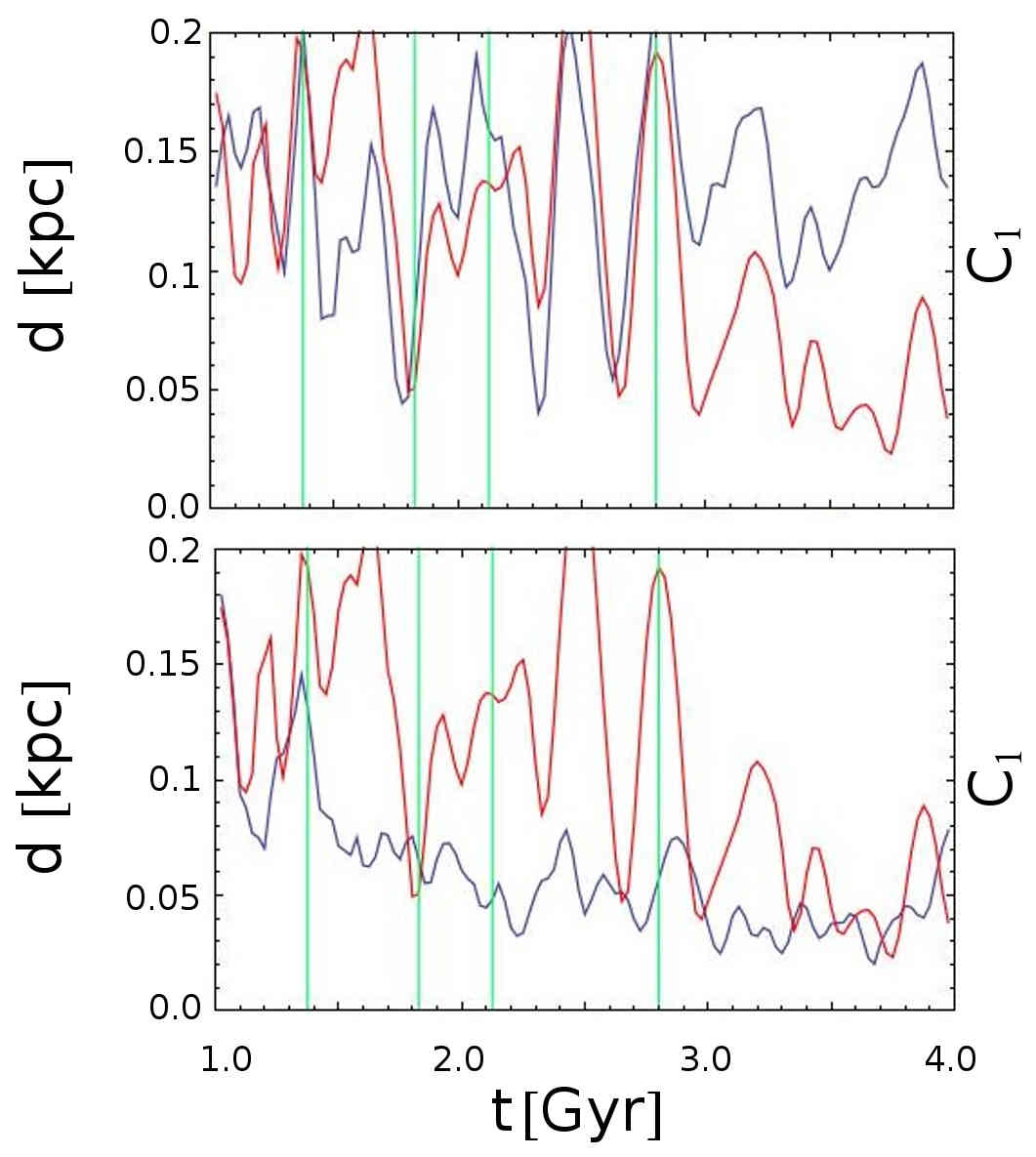}
\caption{\small Top: evolution of the disc-halo displacement $d$
(red, values as in left vertical axis), compared to the mean $C_1$
from three values at $R=3,4$ and $5$ kpc (blue, values as in
right vertical axis). The cyan vertical lines indicate the times
$t_1$ to $t_4$ (same as in Fig.  \ref{fig:hadscorb}).  Bottom: same
as top but for the mean $C_1$ from three values at the radii $R=8,9$
and $10$ kpc.}
\label{fig:distm1}
\end{figure}

The relative disc-halo orbit shown in Fig.~\ref{fig:hadscorb} is quite
small compared to the system's scale.  Fig.~\ref{fig:distm1} shows the 
variations of $d(t)=|\mathbf{\Delta r}_{hd}(t)|$ from $t=1$ Gyr to $4$ Gyr. 
This interval can be split into sub-intervals representing circumcentric or 
epicyclic parts of the disc-halo orbit. In both cases, the distance $d(t)$ 
presents an oscillatory behavior, leading to a minimum distance close to 
$d_{min}\approx 0$ and maximum distance $d_{max}\approx 0.25$ kpc. Despite 
the small size of the disc-halo orbit, the displacement of the halo-bulge 
component with respect to the disc causes a significant $m=1$ perturbation in
the disc. Using a simple model (Appendix I) we find that the amplitude of the 
$m=1$ perturbation scales with the separation as ${\cal O}(d/R_d)$ at 
distances up to $R\sim R_d$, where $R_d$ is the disc's exponential scale 
length. With $d$ as small as $\sim 0.2-0.3$ kpc and $R_d=3$ kpc we find an 
induced $m=1$ perturbation $~0.1$ (see Fig.\ref{fig:habgtide})). Figure 
\ref{fig:distm1} shows direct evidence of the correlation between variations 
of $d$ and of the relative amplitude $C_1$ in the disc. The correlation is 
strong in a domain $3\leq R[\mbox{kpc}]\leq 5$ (top panel), which contains 
the bar beyond its ILR, and connects maxima and minima of $d$ with maxima 
and minima of $C_1$ both in the circumcentric and epicyclic parts of the 
disc-halo orbit. Some change in the mean levels of both $d$ and $C_1$ is 
observed at every orbital switch.  The correlation is weaker at larger 
radii (bottom panel). Thus, the displacement affects the disc's inner part.

\begin{figure}
\centering
\includegraphics[scale=0.18]{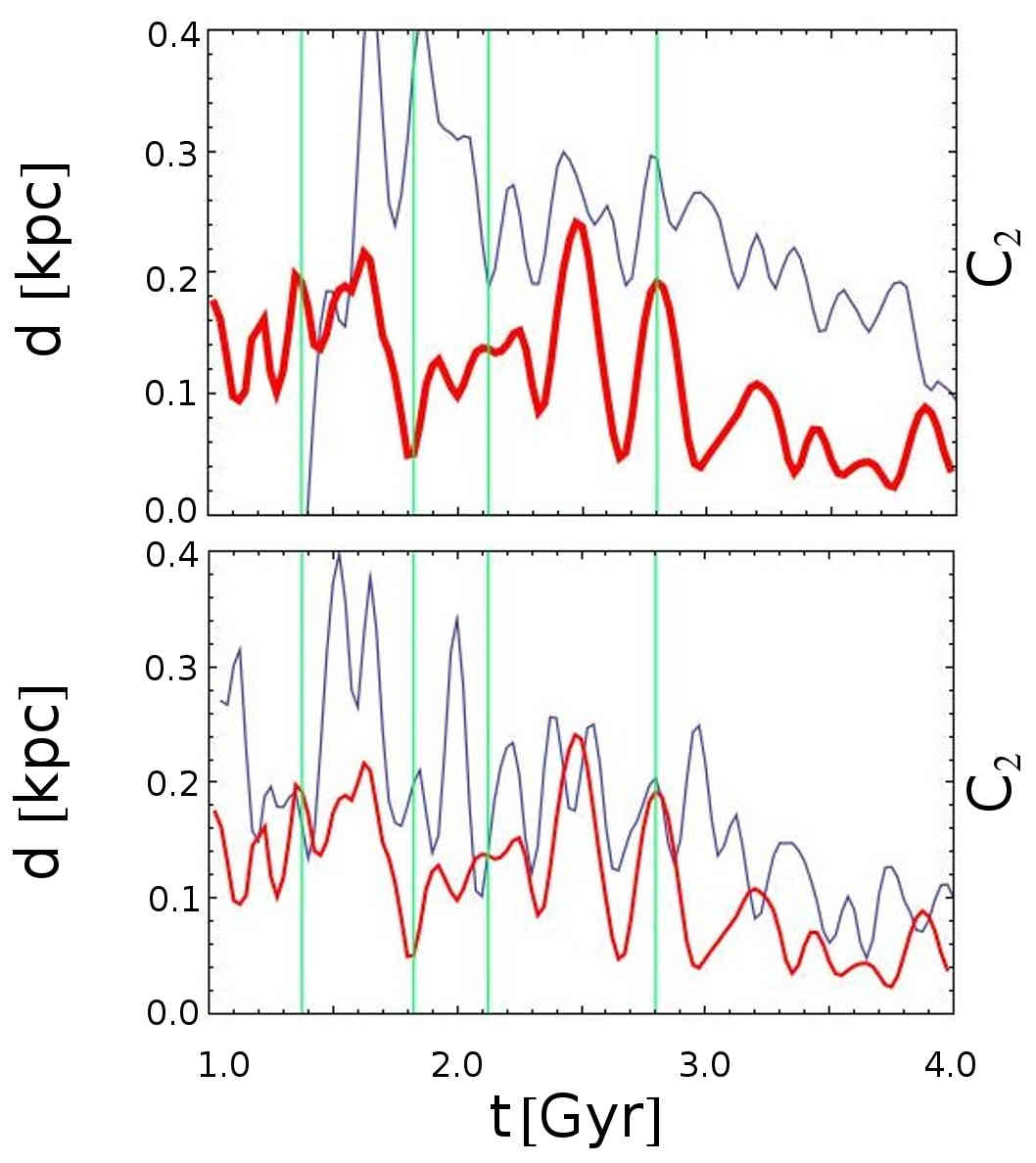}
\caption{\small Same as in \ref{fig:distm1}, but for the mean $C_2$ (blue) 
at the indicated distances.}
\label{fig:distm2}
\end{figure}

The displacement causes also a {\it direct} $m=2$ effect whose size 
is of order ${\cal O}\left((d/R_d)^2\right)$, hence much smaller than the 
$m=1$ effect. However, the {\it orbital} response of disc particles to the
$m=1$ perturbation affects mostly particles inside the bar. Due to 
chaos inside co-rotation (Fig.~\ref{fig:chaos}),
even small potential fluctuations of any harmonics largely influence
the particles' orbits. In particular, particles coming closer to
the manifolds can abandon co-rotation in the direction dictated by the
manifolds. This leads to an {\it indirect} $m=2$ effect. The size of
the effect can be estimated by the correlations between $d$ and the
disc's $C_2$. Figure~\ref{fig:distm2} shows that $d$ and $C_2$ tend to
switch from anti-correlated to correlated when the disc-halo relative
orbit switches from epicyclic to circumcentric. Uncertainties in this
figure obstruct a clear conclusion. As additional evidence,
Figs.~\ref{fig:inburst} and~\ref{fig:outburst} show a significant rise
of the $m=1$ (and also $m=3$) component in the bar at the initial
phase of the incidents of both inner and outer origin.  A similar rise
is observed in all major incidents.

\subsection{Manifold spirals and the evolution of non-axisymmetric 
patterns in the disc}
It was argued above that the manifolds act as drivers of the orbital 
evolution for particles in chaotic orbits pushed to move away from co-rotation. 
This effect is pronounced at major incidents of non-axisymmetric activity, 
but the role of manifolds as drivers of the orbital evolution should hold 
independently of the growth or decay in amplitude of such incidents. We now 
compare the evolution of non-axisymmetric patterns in the disc beyond the 
bar with the evolution of the patterns created by the manifolds, {\it both 
in minima and maxima of the non-axisymmetric activity}. The comparison is 
in terms of the shapes of these patterns.

\begin{figure*}
\centering
\includegraphics[scale=0.24]{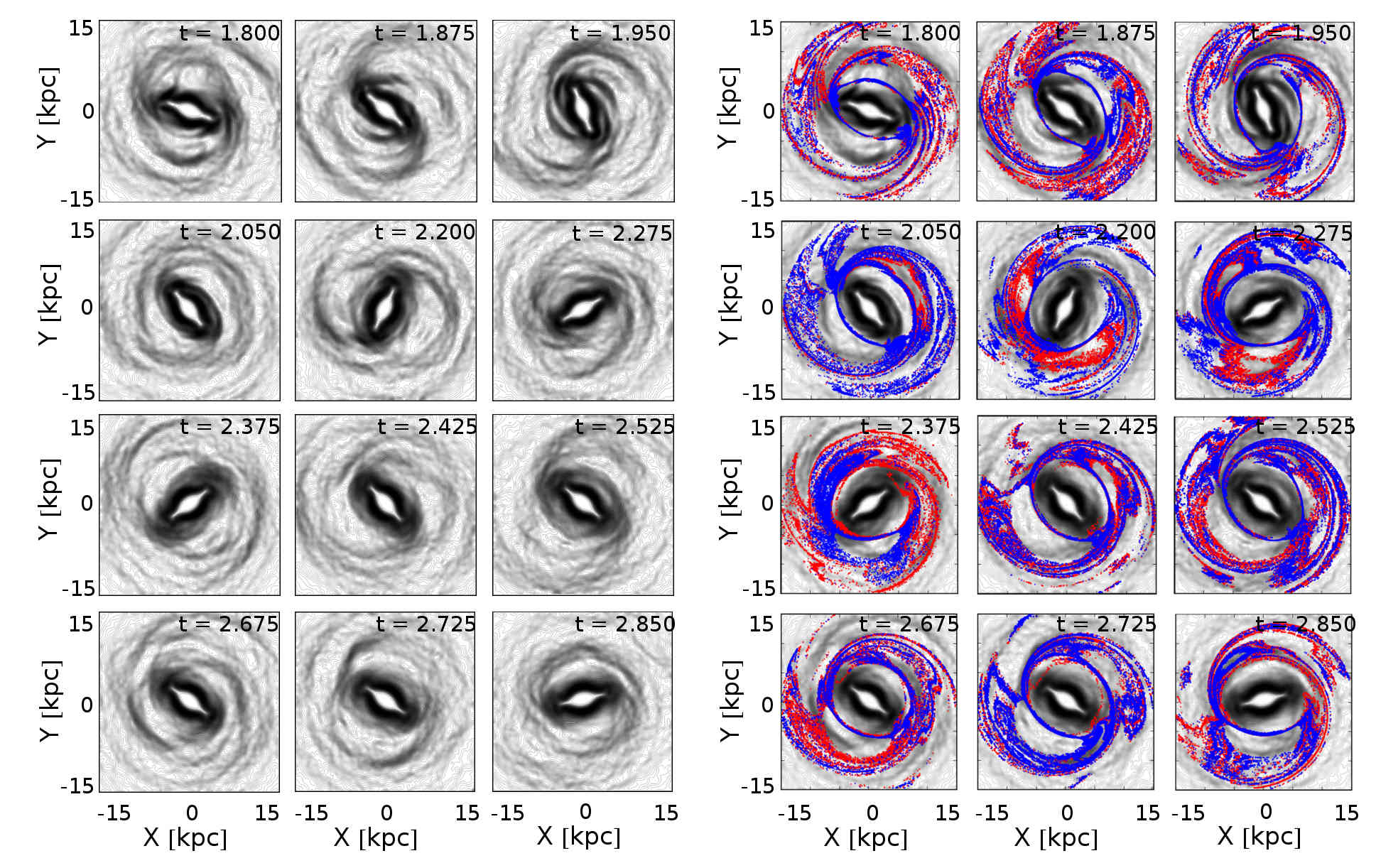}
\caption{\small Left: Sobel-Feldman images of the disc in twelve snapshots 
corresponding to both maxima and minima of $C_2[9\mbox{kpc}]$ (see text). 
Right: the apocentric manifolds for the same snapshots (colors same as in 
Fig.\ref{fig:manall065}), superposed to the Sobel-Feldman images.} 
\label{fig:edgeman}
\end{figure*}
In Fig.~\ref{fig:m12evolve}, nearly in-phase oscillations of the
amplitude $C_2$ occur in the outer parts of the disc. We focus on such
oscillations in the time interval $1.8<t[\mbox{Gyr}]<2.8$ (major
incidents of inner origin, connected with spiral patterns, do not
appear beyond this time). To locate more precisely the times of
occurence of such incidents, we choose a reference radius $R=9$ kpc
which is between the bar's CR and OLR throught the simulation. Six 
consecutive local maxima and six minima of $C_2[9\mbox{kpc}]$ occur 
at the times
\begin{eqnarray}
t^{(max,9\mbox{\small kpc})}[\mbox{Gyr}]
&=&1.8, 1.95, 2.2, 2.375, 2.525, 2.725
\\
t^{(min,9\mbox{\small kpc})}[\mbox{Gyr}]
&=&1.875, 2.05, 2.275, 2.425, 2.625, 2.85~~.\nonumber
\end{eqnarray}
Figure~\ref{fig:edgeman}, left, shows the Sobel-Feldman images of the 
disc in the above times, alternating between a time of maximum and a 
time of minimum.  In a timescale as short as one bar's period one finds 
a strong variability of the patterns identified in the disc beyond 
the bar. Note that the Sobel-Feldman algorithm allows to recover patterns 
which are quite fuzzy in simple surface-density processed images of the 
disc. In particular, some form of spiral activity is identifiable in all 
snapshots of Fig.~\ref{fig:edgeman}, although the morphology of these 
spirals (number of arms, pitch angle, radial extent, etc) rapidly change. 
Besides the main spirals, other secondary features appear and disappear 
in succession, as, for example, rings (e.g. at $t=1.8$ and $t=2.375$ Gyr) 
and bifurcations of more spirals (e.g. at $t=1.95$ Gyr). 
Figure~\ref{fig:edgeman}, right, shows the same images on top of which the 
apocentric manifolds are superposed. An overall comparison shows that the 
manifolds and Sobel-Feldman deduced patterns exhibit a similar level of 
complexity (forming, in the case of the manifolds, a `chaotic 
tangle'~\cite{wig1990}; see also \cite{vogetal2006a}). Also, they agree 
in shape and orientation to a large extent. In more detail, however, the 
level of agreement varies between snapshots. A (rather subjective) visual 
comparison, distinguishes snapshots of better (e.g. $t=1.95, 2.275$ or 
$2.725$ Gyr) or worse agreement (e.g. $t=2.425, 2.625$ Gyr) than average.
The following are some basic remarks regarding this comparison:

- The manifolds' homoclinic lobes account for the systematic appearance 
of features called gaps, bridges and bifurcations in section 3. Such features 
appear in nearly all panels of Fig.~\ref{fig:edgeman}, and show a counterpart 
in the Sobel-Feldman disc images. We note that gaps and bridges can be 
recognized also in several plots of `flux-tube' manifods (see \cite{ath2012}), 
but they are better visualized using the apocentric manifolds.

- In nearly all patterns there are small phase differences ($\leq 5^\circ$ 
between some manifold lobes and the corresponding spirals in the Sobel-Feldman 
image (the largest difference is found for $t=2.375$ Gyr). This might be a 
dynamical phenomenon (disc response has delay with respect to the manifolds), 
but it may also be an artefact of the visualization by the apocentric 
manifolds, since the disc's local density maxima are close to, but do 
not necessarily coincide with the individual orbits' apocenters 
(\cite{tsouetal2008}).

- For computational convenience, we just choose one value of the Jacobi 
constant for the manifold computation. The manifolds retain a rather robust 
pattern against changes of the Jacobi energy (\cite{tsouetal2008}), but small 
variations in shape are allowed. Most notably, besides the manifolds of the 
PL1 or PL2 orbits, the chaotic tangle is supported by the manifolds of 
{\it all} families of unstable periodic orbits in the corotation region. 
Superposed one on top of the other, these manifolds form amore robust 
spiral pattern called `manifold coalescence' in \cite{tsouetal2008}. 
The manifold coallesence is a better representation of chaotic dynamics 
in the co-rotation zone than the manifolds of any single family.  
Nevertheless, its computation for many snapshots is a voluminous work 
extending outside our present scope.

- Finally, despite the clear correlation between manifolds and
Sobel-Feldman detected patterns, not all patterns in the disc need
necessarily be linked to the manifolds. Structures such as outer rings
or slowly rotating outer spirals may not be supported by chaotic flows, 
but instead, by the regular flows associated with standard density wave 
theory (\cite{booetal2005}; \cite{stru2015}).

\subsection{Consistency with multiple pattern speeds}
\begin{figure}
\centering
\includegraphics[scale=0.1]{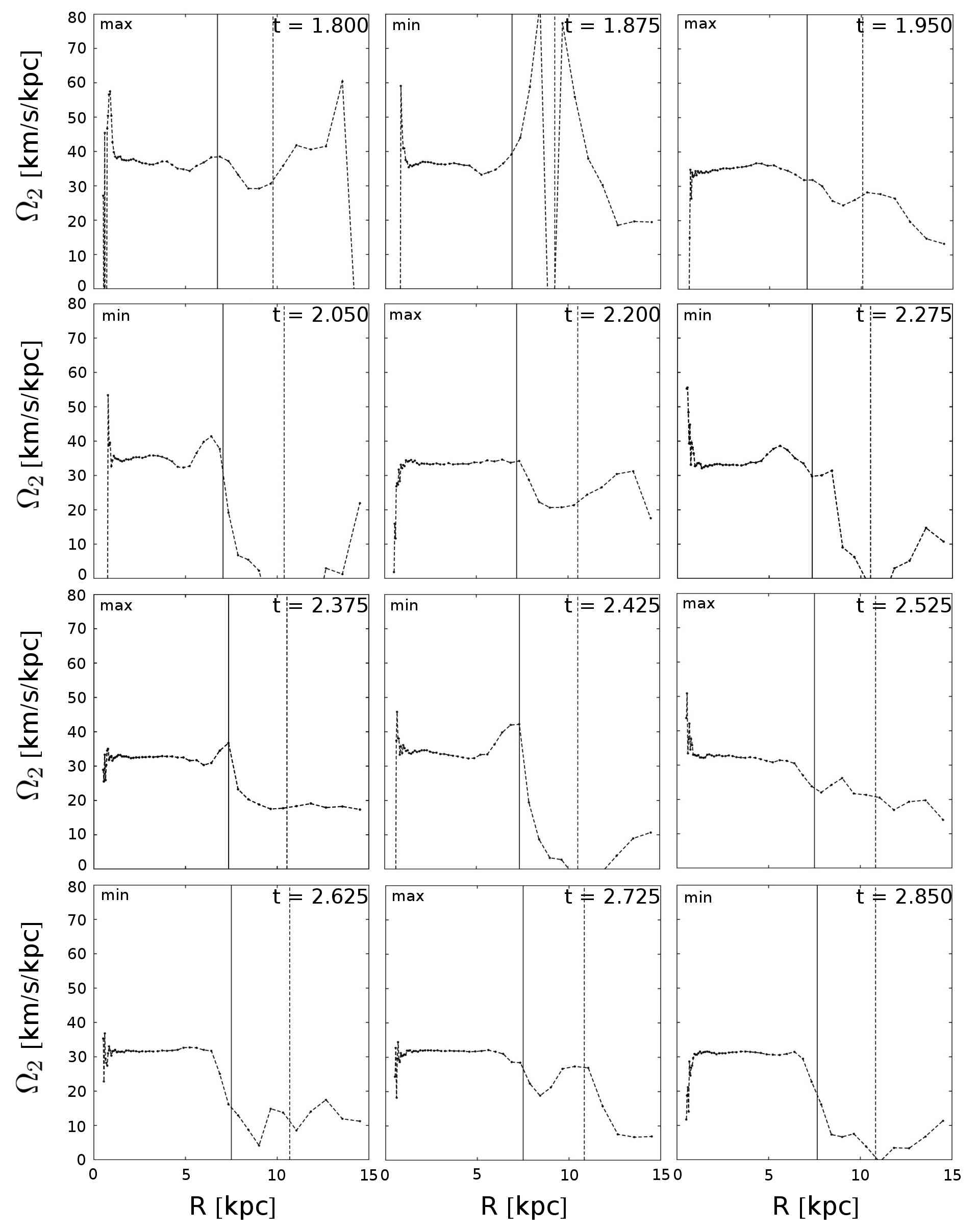}
\caption{\small Profiles of the pattern speed $\Omega_2$ as a function 
of the radius $R$ from the disc center at the same snapshots as in 
Fig. \ref{fig:edgeman}.  The inner plateau at distances between $2$ kpc 
and $4$ kpc determines the bar's pattern speed. The two vertical lines 
mark the position of CR and OLR. Snapshots close to a local maximum or 
minimum of the non-axisymmetric activity beyond the bar (parameterized 
by $C_2[9\mbox{kpc}]$; see text) are marked by `max' and `min' respectively.}
\label{fig:omepatime}
\end{figure}
The fact that multiple pattern speeds are detected in observations and
simulations is considered as an argument against manifold-supported spirals 
(see, for example, \cite{speroo2016}). However, as argued above, measured 
pattern speeds can be affected both by the change in their morphology and 
by the transport of material along invariant manifolds (\cite{ath2012}).

In order to quantify the evolution of pattern speeds, we produce
profiles $\Omega_2(R)$ as in Fig. \ref{fig:omep65}, but for different
snapshots.  Figure~\ref{fig:omepatime} shows these profiles for the
same snapshots as in Fig.~\ref{fig:edgeman}, corresponding to the
consecutive maxima and minima of the $C_2[9\mbox{kpc}]$ amplitude. In
all these snapshots, the inner plateau of nearly constant value
$\Omega_2(R)$ marks the pattern speed of the bar. This plateau extends
up to a distance equal to $\sim 0.6-0.8$ the bar's CR. However,
important fluctuations of the pattern speed $\Omega_2$ beyond CR are
observed at all snapshots after $t=1.8$ Gyr. From the fluctuating
part of the $\Omega_2(R)$ profile between CR and the OLR, the
following evolution pattern is identified: at the six snapshots of
maximum of $C_2[9\mbox{kpc}]$ (`max' in Fig.~\ref{fig:omepatime}),
after a possible hump near corotation $\Omega_2$ starts decreasing,
but tends to stabilize again to a value $\Omega_2<\Omega_{bar}$ as we
approach the OLR. Measured values for this second `plateau' are $\sim
20$ km/sec/kpc, which are a factor $1.5 - 2$ smaller than the (time
decaying) value of $\Omega_{bar}$. Such stabilization disappears at
the minima of $C_2[9\mbox{kpc}]$ (`min' in Fig.~\ref{fig:omepatime}),
where $\Omega_2$ constantly decreases with $R$ beyond CR. Most
notably, the decrease leads to $\Omega_2\approx 0$ at a radius $R$
always close to the OLR.  Decaying profiles of pattern speeds are
reported in observations (e.g \cite{spewes2012};
\cite{speroo2016}). It is of interest to check whether the criterion
of where the decaying curve $\Omega_2(R)$ terminates can be exploited
in real observations for the location of resonances. A systematic
study of this topic is proposed.

On the other hand, Fig. \ref{fig:edgeman} shows no appreciable difference 
in the levels of agreement between manifolds and disc patterns depending 
on whether we are at a maximum or minimum of $C_2[9\mbox{kpc}]$. 
We conclude that the level of agreement is independent of the variability 
of pattern speeds beyond the bar. One may remark, in respect, that multiple 
patterns enhance chaos by the mechanism of resonance overlap 
(\cite{chi1979}; \cite{qui2003}; \cite{minqui2006}) and thus render 
more particles' orbits ruled by manifold dynamics. On the other hand, 
whether or not the manifolds are populated by sufficiently many particles 
to dominate the global patterns in the disc depends on mechanisms able to 
inject new particles in chaotic orbits.  Such mechanisms are distinct 
from the manifolds, as discussed in subsections 4.2 and 4.3.

\subsection{Disc thermalization}
As a final investigation, we examine the time evolution of the disc's
`temperature', i.e., velocity dispersion profile, which is a crucial
factor affecting both the disc's responsiveness to perturbations
as well as the particles' orbits responsiveness to manifold dynamics
(subsection 4.5).  We quantify disc temperature in an annulus of
width $\Delta R$ around the radius $R$ by the radial velocity
dispersion $\sigma_R=\sum\left(V_{R,i}-\mu_R\right)^2$, where
$V_{R,i}$ is the radial velocity of the i-th particle in the annulus
and $\mu_R=\sum_iV_{R,i}$. Similar results hold for the dispersion in
the transverse and vertical velocity components.

\begin{figure}
\centering
\includegraphics[scale=0.35]{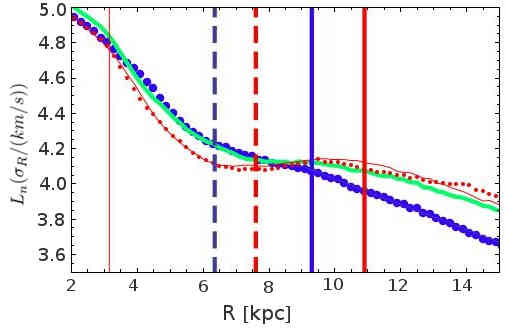}
\caption{\small Profile of the radial velocity dispersion $\sigma_R(R)$ 
at four different times, namely $t=1.625$ Gyr (thick blue with points), 
$t=2$ Gyr (thick green solid), $t=2.5$ Gyr (dotted red) and $t=3$ Gyr (solid 
red). The dashed vertical lines mark the position of CR, and the solid 
vertical lines mark the position of the OLR at $t=1.625$ Gyr (blue) and 
$t=3$ Gyr (red) respectively.  } 
\label{fig:sigmar}
\end{figure}

Figure \ref{fig:sigmar} shows the profile $\sigma_R(R)$ at four
different times, namely $t=1.625$, $t=2$, $t=2.5$ and $t=3$ Gyr. At
$t=1.625$, the dependence of the radial velocity dispersion on $R$ can
be approximated by the union of two exponential profiles, one for the
inner part of the disc (in the interval $R_{in}\leq R\leq R_{CR}$),
with law $\sigma_R\approx \sigma_1\exp(-(R-R_{in})/R_1)$, with
$\sigma_1\approx 150$ Km/sec, $R_1\approx 6$ kpc, $R_{in}\approx 2$
kpc, and another for the outer part of the disc ($R>R_{CR}$), with law
$\sigma_R\approx \sigma_2\exp(-(R-R_{CR})/R_2)$, with $\sigma_2\approx
70$ Km/sec, $R_2\approx 15$ kpc.
Three more curves in Fig.~\ref{fig:sigmar}, corresponding to $t=2,
2.5$ and $3$ Gyr, show how the profile $\sigma_R$ versus $R$ evolves
in time.  At radii beyond the last position of the OLR ($R>11$ kpc),
the velocity dispersion increases in time by a factor between $1.1$
and $1.2$ in a time interval $\sim 1.5$ Gyr, and the corresponding
exponential profile becomes less steep. We notice, however, that the
profile becomes nearly horizontal, leading to an `isothermalization'
(constant velocity dispersion) in a domain of the disc roughly between
$R=6$ kpc and $R=11$ kpc.  As seen in Fig. \ref{fig:sigmar}, the inner
limit of this domain nearly coincides with the innermost position of
CR (at $t=1.625$), while the outer limit nearly coincides with the
outermost position of the OLR (at $t=3$ Gyr) in the considered time
interval. One remarks that, as they move outward, at least one of
these resonances visits part of the interval $6.5<R[kpc]<11$, in which
the isothermalization takes place. Physically, the domain
$R_{CR}<R<R_{OLR}$ is where chaotic motions completely dominate the
dynamics. The corresponding particles belong to the well known `hot
population' (\cite{spasel1987}), whose orbits span the whole domain
while recurrently entering also inside corotation. As a result, the
strongly chaotic dynamics brings about an equalization of the
velocity dispersion in the whole domain. On the other hand, we observe
that the velocity dispersion profile remains practically invariant at
distances smaller than the initial position of the ILR (at $\approx 3$
kpc). In fact, this resonance appears to act as a barrier for `heat'
(random kinetic energy) transfer across the disc.

The key remark, regarding the above evolution, is that although incidents 
of non-axisymmetric activity (in particular spirals) heat the outer parts 
of the disc, isothermalization in the domain scanned by the CR and OLR 
resonances implies that a part of the disc between CR and OLR becomes 
{\it cooler} with time. Physically, particles in the outer parts 
of the bar migrate outwards, carrying with them kinetic energy in the 
form of random motions. In Fig.~\ref{fig:sigmar}, the initial and final 
profiles $\sigma_R(R)$ intersect at a radius $R_c\approx 8.5$ kpc. 
The bar's co-rotation at the time of the initial profile is at 
$R_{CR} \approx 6.4$ kpc, while it shifts to $R'_{CR}\approx 7.5$ kpc 
at the final time. Estimating by $R_{CR}<R<R_c$ the disc domain where 
manifolds rule the response of particles' orbits to external perturbations, 
this domain has a width $\sim 2.1$ kpc at the initial time $t=1.6$ Gyr, 
which is limited to $\sim 1$ kpc at $t=3$ Gyr, while this domain is 
expected to shrink further to negligible widths as the bar's co-rotation 
keeps moving outwards.

\section{Discussion and conclusions}

The origin and longevity of spiral structure in disc galaxies is still matter 
of intense debate. As new observational data and more accurate simulations 
come to surface, we gradually arrive at appreciating the potential 
complexity of the mechanisms which generate and maintain spiral structure. 
The co-existence and possible non-linear coupling of multiple patterns  
(see references in introduction) hints towards the chaotic nature of spiral 
arms, in particular when a strong bar is present. To the extent that spirals 
are bar driven, they are affected in a non-trivial way by the bar's secular 
evolution. Within this context, the main contribution of the manifold theory 
is to describe {\it what should be the expected form of the bar-driven spiral 
mode when the disc's region between the bar's Corotation and Outer Lindblad 
Resonance is largely chaotic}. It should be stressed that the manifold theory 
poses no requirement that chaos originates exclusively from the bar. Nonlinear 
interaction of the bar mode with additional patterns beyond co-rotation 
actually enhances chaos (see references in text).  Furthermore, it is a basic 
rule of dynamics that the unstable manifolds of one periodic orbit can 
intersect neither themselves nor the unstable manifolds of any other periodic 
orbit of equal Jacobi energy. Thus, all these manifolds have to `coalesce' 
in nearly parallel directions, thus enhancing chaotic spirals. 

The manifolds emanating from the region of the bar's $L_1$ and $L_2$ points 
provide the simplest representation of these chaotic outflows. In the present 
paper we give evidence that these manifolds provide a dynamical skeleton in 
phase space, or, the dynamical avenues to be followed by new particles 
injected in the domain between CR and OLR at reccurent `incidents' of 
non-axisymmetric activity. We gave a characterization of such incidents as 
of i) inner, or ii) outer origin, depending on whether the spectral analysis 
shows a wave originating inside CR and propagating outwards (in (i)), 
or originating outside CR, moving initially inwards, then being reflected 
at CR, and then moving outwards (in (ii)). Manifold spirals are 
connected with incidents of inner origin. Pattern detection algorithms such 
as Sobel-Feldman allow to detect the co-existence of various patterns, 
both at maxima and minima of the $m=2$ amplitude beyond the bar. 
We interpret the importance of manifolds as follows: whatever causes these 
incidents, at every incident the particles' orbits (and in particular 
chaotic ones) are perturbed. Then, new particles injected in the CR zone 
tend to follow and populate the manifolds to a large extent, according 
to general rules of dynamics. 

We argued that the continuous change of the form of the manifolds, as
well as the motion, along the manifolds, of the matter which populates
them, allows to reconcile manifold spirals with the multiplicity
(and variability) of pattern speeds beyond the bar. A basic behavior is 
found to govern the evolution of the radial profile of the pattern frequency 
$\Omega_2(R)$ beyond the bar. At maxima of the non-axisymmetric activity 
$\Omega_2(R)$ tends to form a second plateau, indicating a second pattern 
speed outside CR, well distinct from $\Omega_{bar}$. At minima, the second 
plateau disappears, giving its place, to a shear-indicating decaying profile
of the curve $\Omega_2(R)$. Remarkably, in the latter case, $\Omega_2(R)$ 
terminates at a value $\Omega_2\sim 0$ always near $R=R_{OLR}$. Five full 
cycles of this behavior are seen in our simulation, in a period of 
$\sim 1$ Gyr (see Fig.~\ref{fig:omepatime}), leading to an approximate 
period of $\sim 0.2$ Gyr. This is in rough resonance with the bar's period, 
but with large uncertainties in the numbers. On the other hand, since the
comparison between manifolds and Sobel-Feldman-recognized patterns in
the disc (Fig.~\ref{fig:edgeman}) shows no appreciable differences
between snapshots of minima and maxima of the non-axisymmetric
activity, we argued that the degree to which manifolds are able to
dictate the dynamics outside CR seems to be rather independent of the
strength of any additional pattern in the disc.

In addition to well known mechanisms, we discussed a new mechanism which 
seems to play a key role in triggering new incidents of non-axisymmetric 
activity: this is the relative halo-disc orbit which results from the 
off-centering effect: whenever particles are ejected away from the disc 
through manifold spirals, the disc rebounces, leading to a relative orbit 
between the disc and the remaining spheroid (halo+bulge) component of the 
galaxy. We argue that even moderate off-centerings, of order $d/R_d\sim 0.1$, 
where $d$ is the off-center displacement and $R_d$ the disc's exponential 
scale length, are able to induce appreciable $m=1$ tides on the disc 
(of order ${\cal O}(d/R_d)$). The corelation between $d$ and the disc 
$m=1$ response is clearly seen in the disc's spectral analysis. The way 
the $m=1$ tide affects the $m=2$ response beyond the bar appears to be a 
complex nonlinear phenomenon for which we have no precise analysis at present. 
Yet, the fact that one phenomenon influences the other is clearly seen in 
our simulation, as detailed in subsections 4.2 to 4.4. 

As a final remark, we discussed the `thermal' evolution of the disc, i.e., 
the way in which the radial profile of the velocity dispersion appears to be 
influenced in time due to non-axisymmetric activity. Owing, again, to the 
large degree of chaos between CR and OLR, we argued that the gradual outward 
shift of both the CR and OLR radii as the bar slows down, causes part of the 
disc, beyond the end of the bar, to acquire a nearly constant radial velocity 
dispersion. In fact, this `isothermalization' causes a part of the disc,  
starting from inside the bar and ending at a point midway between CR and 
OLR to cool down. We interpret this effect as a hint 
that chaotic populations of particles gradually migrate outwards, carrying 
with them kinetic energy in the form of randomly oriented motions. 
Regarding, however, the responsiveness of the disc to manifold dynamics, 
we provide a heuristic argument showing that good levels of responsiveness 
should be limited in a domain which shrinks in time, as the CR radius moves 
outwards, while the radius beyond which the disc gets hotter in time is 
nearly fixed. 

\section*{Acknowledgements}
We thank the reviewer, Prof. J. Binney, for his detailed report 
with many suggestions for improvement. We thank Prof. E. Athanassoula 
for a very careful reading of the manuscript, with many critical comments, 
as well as for enlightening discussions on manifold dynamics. We also 
thank Prof. P. Patsis for useful discussions. The authors acknowledge the Greek 
Research and Technology Network (GRNET) for the provision of the National 
HPC facility ARIS under Project PR004033-ScaleSciComp II.  R.I. Paez was 
funded by the Project 200/854 of the Research Committee of the Academy of Athens
and the ERC Project 677793 StableChaoticPlanetM, in non overlapping periods.
The potential interpolation algorithm and manifold computations were 
developed as part of K. Zouloumi's MSc thesis (Department of Physics, 
University of Athens, 2017).



\appendix
\section{Disc-halo relative displacement and $m=1$ perturbation 
on the disc}

An estimate of the strength of the $m=1$ perturbation on the disc
plane due to the disc-halo relative displacement can be obtained as
follows: Consider a simplified model with two constituents, namely a
razor-thin disc with exponential surface density profile
$\Sigma(R)=\Sigma_0\exp(-R/R_d)$, truncated at an inner radius $R_1$,
and a spheroidal component (e.g. the combined halo-bulge system)
represented by the spherical potential $V_0(r)$.  Let O, the center of the 
disc, be the origin of the co-ordinates. We assume the center C of the 
spherical component to be displaced with respect to O, while remaining 
in the disc plane (as indicated by the simulations, see section 4). 
Let $\mathbf{d}$ be the displacement vector connecting O to C.  
Cartesian co-ordinates $(x,y)$ are defined such that the $x-$axis is 
parallel to $\mathbf{d}$.

The potential $V_0(r)$ expressed in polar co-ordinates $(R,\phi)$, with 
$x=R\cos\phi$, $y=R\sin\phi$, reads
\begin{equation}\label{eq:pot0}
\Phi(R,\phi)=V\left((R^2+d^2-2Rd\cos\phi\right)^{1/2})~~,
\end{equation}
where $d=|\mathbf{d}|$. For small displacements ($d<<R_1<R$), the 
potential (\ref{eq:pot0}) can be developed in powers of the ratio $d/R$. 
Up to first degree in $d/R$ we have:
\begin{equation}\label{eq:pot0exp}
\Phi(R,\phi)=V(R)-V'(R)R\left({d\over R}\right)\cos\phi+
{\cal O}\left(\left({d\over R}\right)^2\right)
\end{equation}
As obvious from Eq.~(\ref{eq:pot0exp}), the disc-halo relative
displacement induces a $m=1$ component of the potential generated by
the spherical component on the disc plane. Specifically,
$\Phi(R,\phi)=\Phi_0(R)+\Phi_1(R)\cos\phi+...$, with
\begin{equation}
  \Phi_0(R)=V(R), ~~~ \Phi_1(R)=-V'(R) R\left({d\over R}\right)~~.
\end{equation}
Higher order terms depending on the powers $(d/R)^m$, $m=2,3,\ldots$ 
are in general negligible. The residual $m=1$ gravitational field as measured 
by an observer constantly at the disc's center of mass is found after 
subtracting a non-inertial term generated by the disc's center-of-mass 
acceleration:
\begin{equation}
a_d = \left({d\over R_d}\right)
\frac{\int_{R_1}^\infty \left(V''(R)R+V'(R)\right) R_d e^{-R/R_d} dr}
{\int_{R_1}^\infty e^{-R/R_d} r dr}
\end{equation}
The residual potential is then given by
\begin{equation}\label{eq:potidal}
\Phi_{res} = \Phi_d(R)+\Phi_0(R) +\Phi_1(R)\cos\phi - a_d R\cos\phi + ...
\end{equation}
The importance of the term due to $a_d$ can be estimated for particular 
choices of potentials $V(R)$. For a Keplerian potential $V(R)\sim -GM/R$ 
we have
\begin{equation}
a_d \sim - c {GM\over R_d^2}\left({d\over R_d}\right), ~~~
c={1\over 2}
\frac{\int_{R_1/R_d}^\infty e^{-\xi} \xi^{-2}d\xi}
{\int_{R_1/R_d}^\infty e^{-\xi} \xi d\xi}
\end{equation}
We find $c\sim 0.1$ for $R_1/R_d\sim 1$. Thus, $a_d$ is a factor 
$\sim 0.1(d/R_d)$ smaller than the axisymmetric force produced by the 
spheroidal component on the disc at the distance $R\sim R_d$. We notice also 
that $a_d$ has a negative sign, i.e., as a result of their mutual displacement, 
the disc center of mass is {\it repelled} by the spheroidal component. 

\begin{figure}
\centering
\includegraphics[scale=0.19]{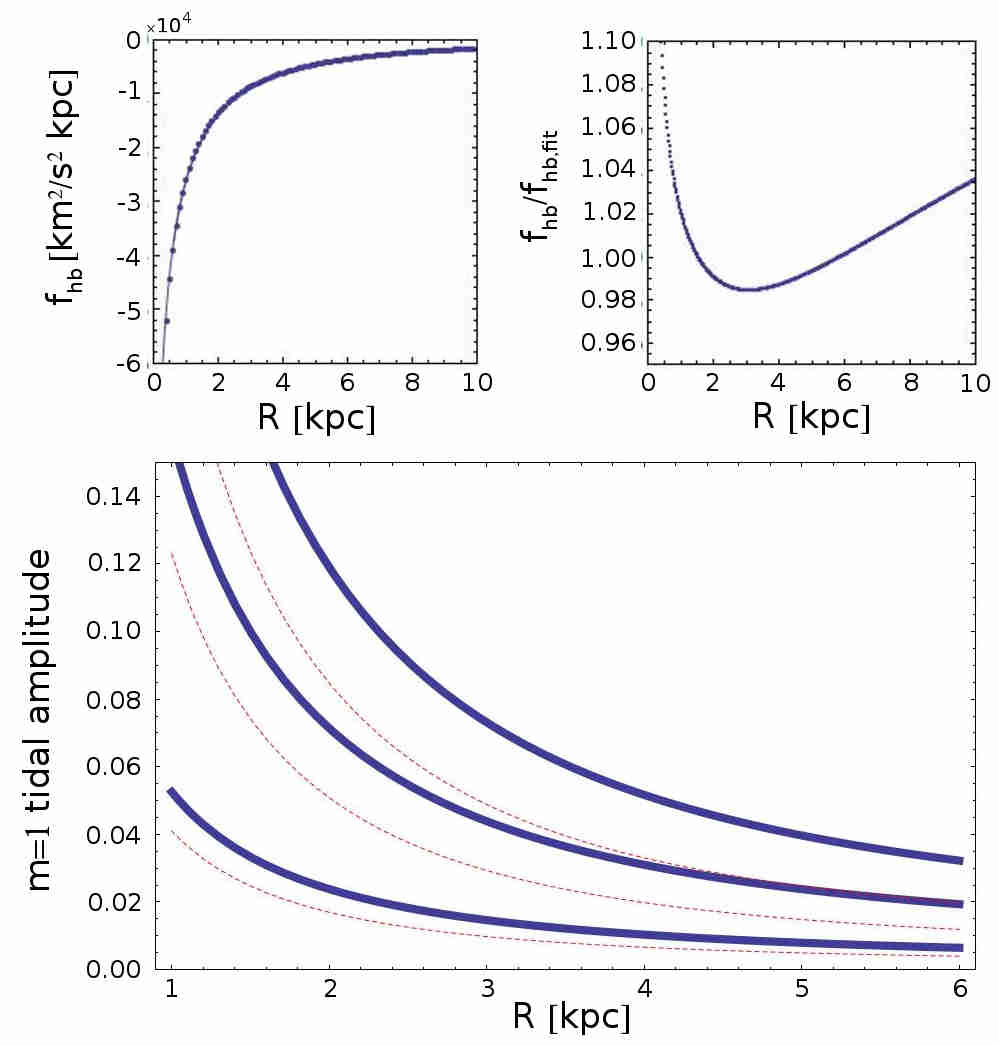}
\caption{\small Top left: comparison between the halo-bulge combined central force 
per unit mass $f_{hb}$ as a function of the radius $R$ and the force $f_{hb,fit}$ 
from the double-Hernquist fitting formula (Eq:\ref{eq:hernfit}) with the adopted 
parameters (see text). Top right: the ratio $f_{hb}/f_{hb,fit}$ as a function of 
$R$. Bottom: the absolute ratio $|(f_1^2+g_1^2)^{1/2}/f_0|$ of the $m=1$ over 
axisymmetric force when the halo-bulge displacement is $d=0.25$ kpc,$d=0.15$ kpc 
or $d=0.05$ kpc (top, middle and bottom pairs of curves respectively). The solid 
blue curves correspond to $f_0$ being computed using the halo-bulge pair only, 
while for the red dashed curves $f_0$ includes also the disc axisymmetric force.} 
\label{fig:habgtide}
\end{figure}

As an application, we will compute the relative amplitude of the $m=1$
component of the residual force in a potential $V(R)$ representing the sum 
of the bulge and halo components of our N-body simulation. The bulge is 
given by a Sersic law with scale length $r_b=1$ kpc, Sersic index $n=3.5$ 
and total mass $5\times 10^9M_\odot$. The halo is given by a double power-law 
density profile $\rho(r) = \rho_0(r/r_h)^{-\alpha_h}(1+r/r_h)^{\beta_h-\alpha_h}$ 
with $\alpha_h=1.3$, $\beta_h=3.5$, and $\rho_0=2.016\times 10^8$
$M_{\odot}/kpc^3$. We fit the total potential using a double Hernquist
potential:
\begin{equation}\label{eq:hernfit}
V_{fit}(R) = V_0-\sum_{i=1}^2 {GM_i\over R+a_i}
\end{equation}
with $V_0=2500 Km^2/sec^2$, $a_1=0.5$ kpc, $a_2=3$ kpc, and 
$M_1=5\times 10^9 M_{\odot}$, $M_2=6.2\times 10^{10}M_{\odot}$. The top panels 
in Fig. \ref{fig:habgtide} show the fitting of the force produced by the spheroid 
by using the fitting formula (\ref{eq:hernfit}) instead of the exact potential 
and forces of the adopted halo-bulge model. Finally, we consider an exponential 
disc with $R_d=3$ kpc, and $\Sigma_0=M_d/(2\pi R_d^2)$, with $M_d=5\times 
10^{10}M_{\odot}$. With the above formulas, the residual potential 
(\ref{eq:potidal}) is given by:
\begin{eqnarray}\label{eq:phiexpd}
\Phi_d(R) = -G\Sigma_0\pi R \times~~~~~~~~~~~~~~~~~~~~~~~\nonumber\\
\Bigg(I_0\left({R\over 2R_d}\right)K_1\left({R\over 2R_d}\right)
-I_1\left({R\over 2R_d}\right)K_0\left({R\over 2R_d}\right)\Bigg)
\end{eqnarray}
\begin{eqnarray}\label{eq:phiexp0}
\Phi_0(R)=-\sum_{i=1}^2{GM_i\over R+a_i},~~~
\Phi_1(R)=-\sum_{i=1}^2{GM_i d\over (R+a_i)^2}
\end{eqnarray}
\begin{eqnarray}\label{eq:adexp}
a_d = -\sum_{i=1}^2 c_i {GM_i d\over R_d^3},~c_i = {1\over 2}
\frac{\int_{\xi_1}^\infty 
e^{-\xi}\left({\xi-a'_i\over \xi(\xi+a'_i)^3}\right)d\xi}
{\int_{\xi_1}^\infty 
e^{-\xi}\xi d\xi}~~.
\end{eqnarray}
$I_0, I_1, K_0, K_1$ are cylindrical Bessel functions, and $a'i = a_i/R_d$, 
$\xi_1=R_1/R_d$.  Differentiating the potential (\ref{eq:potidal}) obtained 
through the above formulas, we find the residual forces 
$F_{R,res}=-\partial\Phi_{res}/\partial R$, 
$F_{\phi,res}=-(1/R)\partial\Phi_{res}/\partial\phi$ as: 
\begin{eqnarray}\label{eq:fresexp}
F_{R,res}(R,\phi) &=& f_0(R) + f_1(R)\cos\phi,~~~ \\
F_{\phi,res}(R,\phi) &=& g_1(R)\sin\phi~~. \nonumber
\end{eqnarray}
Figure \ref{fig:habgtide} shows the absolute ratio
$|(f_1^2+g_1^2)^{1/2}/f_0|$ for the adopted model parameters, as a
function of the distance $R$ from the disc center, for three values of
the displacement $d=0.05, 0.15$ and $0.25$ kpc. One notices that even
small displacements result in $m=1$ fields of amplitude a few percent
in the inner parts of the disc. In particular, for the largest observed 
displacement in the simulation, $d=0.25$ kpc, the $m=1$ relative amplitude 
exceeds $5\%$ at the disc's inner $3$ kpc, and $10\%$ inside the inner 
$2$ kpc.


\bsp	

\label{lastpage}
\end{document}